\documentclass[fleqn,a4paper]{article}
\usepackage{a4wide}
\usepackage{amsmath}
\usepackage{amssymb}
\usepackage{comment}
\usepackage{hyphenat}
\usepackage{multicol}
\usepackage{graphicx}
\usepackage{amsfonts}
\usepackage{subcaption}
\usepackage{bm}
\usepackage{graphicx}
\usepackage{xcolor}
\usepackage{natbib}
\usepackage[linesnumbered,ruled]{algorithm2e}

\newcommand{\qed}{\hfill$\Box$}

\title{Anomaly and Novelty detection \\for robust semi-supervised learning
}

\author{Andrea Cappozzo \footnote{Department of Statistics and Quantitative Methods, University of Milano-Bicocca, \texttt{a.cappozzo@campus.unimib.it, francesca.greselin@unimib.it}}         \and
        Francesca Greselin\footnotemark[\value{footnote}] \and
        Thomas Brendan Murphy \footnote{School of Mathematics \& Statistics and Insight Research Centre, University College Dublin, \texttt{brendan.murphy@ucd.ie}}
}

\begin{document}
\maketitle

\abstract{
   Three important issues are often encountered in Supervised and Semi-Supervised Classification: class-memberships are unreliable for some training units (label noise), a proportion of observations might depart from the main structure of the data (outliers) and new groups in the test set may have not been encountered earlier in the learning phase (unobserved classes). The present work introduces a robust and adaptive Discriminant Analysis rule, capable of handling situations in which one or more of the afore-mentioned problems occur. Two EM-based classifiers are proposed: the first one that jointly exploits the training and test sets (transductive approach), and the second one that expands the parameter estimate using the test set, to complete the group structure learned from the training set (inductive approach). 
Experiments on synthetic and real data, artificially adulterated, are provided to underline the benefits of the proposed method.}


\section{Introduction} \label{intro}

\begin{figure*}
\begin{subfigure}{.3\textwidth}
  \centering
  \hspace*{-.4cm}\includegraphics[width=1.1\linewidth]{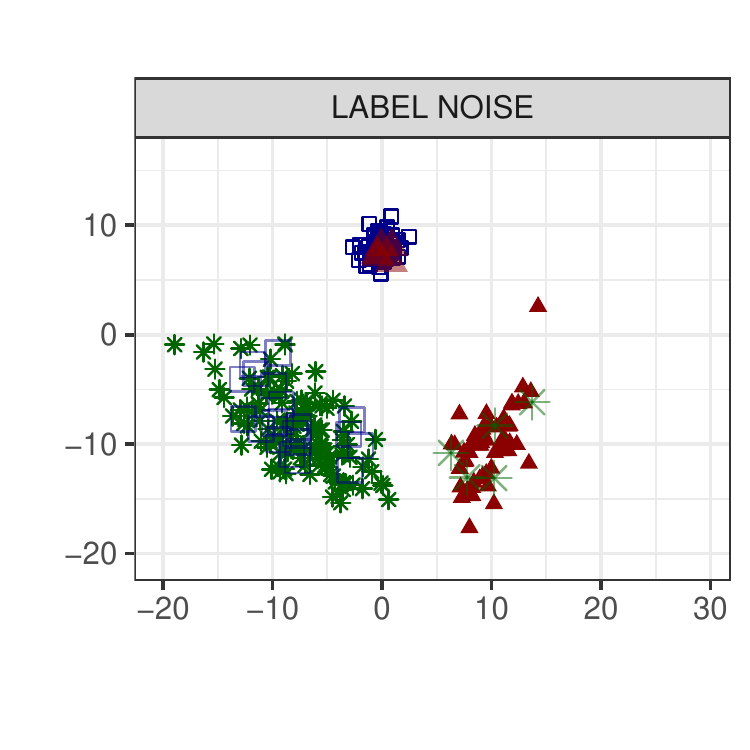}
\end{subfigure}%
\begin{subfigure}{.3\textwidth}
  \centering
\includegraphics[width=1.1\linewidth]{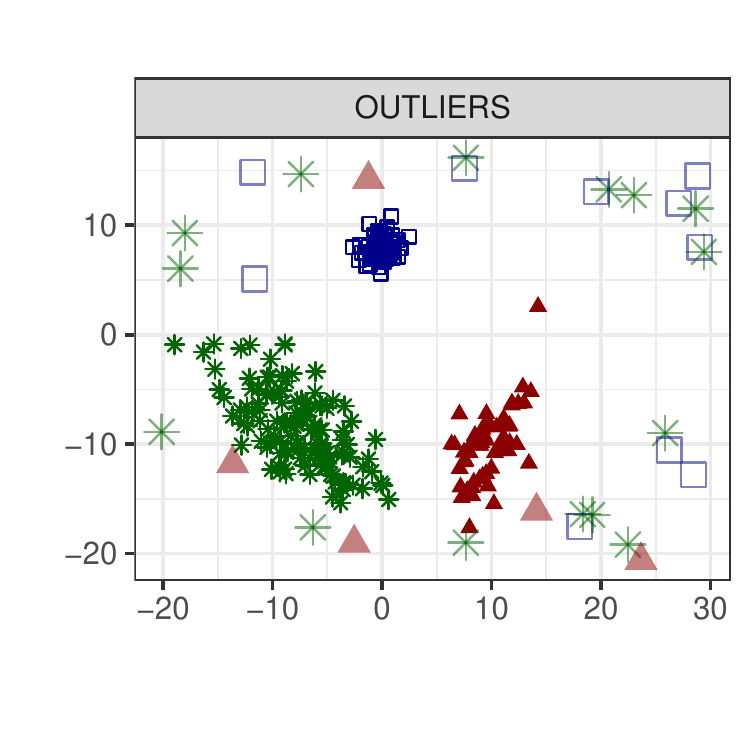}
\end{subfigure}
\begin{subfigure}{.3\textwidth}
  \centering
 \hspace*{.4cm} \includegraphics[width=1.1\linewidth]{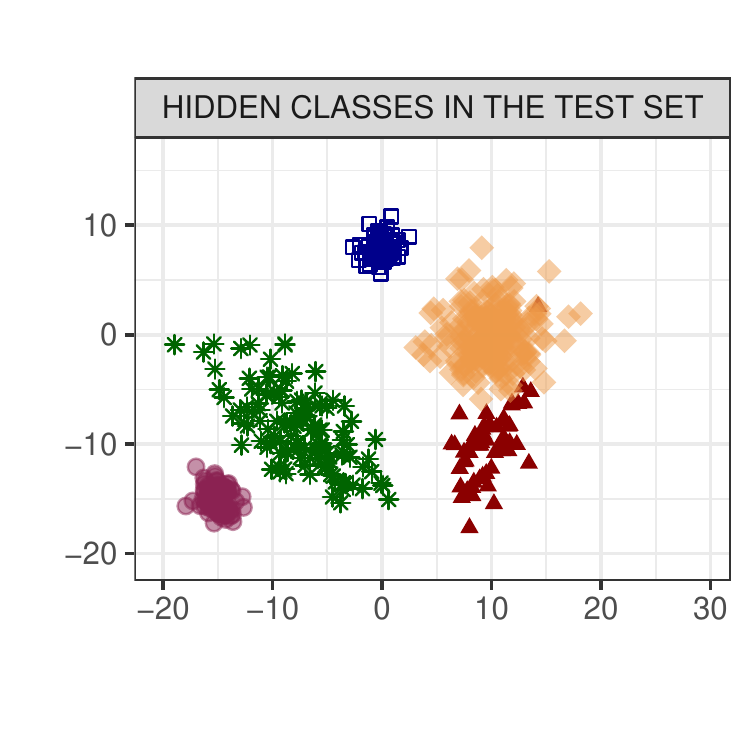}
\end{subfigure}%
\caption{Different classification scenarios in which the training set presents label noise (left panel), outliers (central panel) and in which the test set contains groups not previously encountered in the learning phase (right panel).} 
\label{fig:class_problems}
\end{figure*}

The standard classification framework assumes that a set of outlier-free and correctly labelled units are available for each and every group within the population of interest. Given these strong assumptions, the labelled observations (i.e., the training set) are employed to build a classification rule for assigning unlabelled samples (i.e., the test set) to one of the known groups. However, real-world training set may contain noise, that can adversely impact the classification performances of induced classifiers. Two sources of anomalies may appear:
\begin{itemize}
\item label noise, that is wrongly labelled data, represented in the left panel of Figure \ref{fig:class_problems};
\item feature noise, whenever erroneous measurements are given to some units, as shown in central panel of Figure \ref{fig:class_problems}. 
\end{itemize} 
Moreover, when new data are given to the classifier, extra classes, not observed earlier in the training set, may appear (see right panel of Figure \ref{fig:class_problems}).
Therefore, for a classification method to succeed when the aforementioned assumptions are violated, both anomalies and novelties need to be identified and categorized as such. Since neither anomaly nor novelty detection is universally defined in the literature, we hereafter characterize their meaning in the context of classification methods. 
%

Anomaly detection refers to the problem of finding patterns in data that do not conform to expected behaviour \citep{Chandola2009}. Particularly, we designate as anomalies the noisy units whose presence in the dataset obscures the relationship between the attributes and the class membership \citep{Hickey1996}. Following \cite{Zhu}, we distinguish between attribute and class noise: the former identifies units with unusual values on their predictors (outliers), whilst with the latter we indicate observations with inaccurate class membership (label noise). Examples of methods able to deal with anomalies in classification include Robust Linear Discriminant Analysis \citep{Hawkins1997}, Robust Soft independent modelling of class analogies \citep{VandenBranden2005}, Robust Mixture Discriminant Analysis \citep{Bouveyron2009a}, and, more recently, Robust Updating Classification Rules \citep{Cappozzo2019b}.

Novelty detection is the identification of new or unknown data or signal that a machine learning system is not aware of during training \citep{Markou2003b}. Particularly, in a classification context, we indicate with novelty a group of observations in the test set that displays a common pattern not previously encountered in the training set, and can therefore be identified as a novel or hidden class. From a stochastic viewpoint, this is equivalent to assuming that the probability distribution of the labels differs in the labelled and unlabelled sets, as a result of an unknown sample rejection process. More generally, the difference between the joint distribution of labels and input variables in the training and test sets is denoted as ``dataset shift'' problem: for a thorough description of the topic, the interested reader is referred to \cite{quionero2009dataset}. Examples of methods that are able to deal with novelties in classification include Classifier Instability \citep{Tax1998}, Support Vector Method for novelty detection \citep{Scholkopf2000} and Adaptive Mixture Discriminant Analysis \citep{Bouveyron}. Recently, \cite{fop2018unobserved} extended the latter method to account for unobserved classes and extra variables in high-dimensional discriminant analysis.

The ever-increasing complexity of real-world datasets motivates the development of methods that bridges the advantages of both novelty and anomaly detection classifiers. For instance, human supervision is required in bio-medical applications: this costly and difficult procedure is prone to introduce label noise in the training set, while some less common or yet unknown patterns might be left completely undiscovered. Another example comes from the food authenticity domain: adulterated samples are nothing but wrongly-labeled units in the training set, whilst new and unidentified adulterants may generate unobserved classes that need to be discovered. Also, in food science, the state-of-the-art approach for determining food origin is to employ microbiome analysis as a discriminating signature: wo promising applications for identifying wine provenance and variety are reported in Section \ref{sec:application}. 
 
In the present paper we introduce a novel classification method for situations where class-memberships are unreliable for some training units (label noise), a proportion of observations departs from the main structure of the data (outliers) and new groups in the test set were not encountered earlier in the learning phase (unobserved classes). Our proposal models the unobserved classes as arising from new components of a mixture of multivariate normal densities, and no distributional assumptions are made on the noise component.

The rest of the manuscript is organized as follows. Section 2 briefly describes the adaptive mixture discriminant analysis (AMDA): a model-based classifier capable of detecting several unobserved classes in a new set of unlabelled observations \citep{Bouveyron}. In Section 3, we introduce a robust generalization of the AMDA method employing mixture of Gaussians: 
robustness is achieved via impartial trimming and constraints on the parameter space and adaptive learning is obtained by means of a transductive or inductive EM-based procedure. Model selection is carried out via robust information criteria. Experimental results for evaluating the features of the proposed method are covered in Section 4. Section 5 presents two real data applications, involving the detection of grapes origin and must variety when only a subset of the whole set of classes are known in advance and learning units are not to be entirely trusted. 
Section 6 concludes the paper with some remarks and directions for future research.
\section{Adaptive mixture discriminant analysis} \label{sec:AMDA}
The Adaptive Mixture Discriminant Analysis (AMDA), introduced in \cite{Bouveyron}, is a model-based framework for supervised classification that accounts for the case where some of the test units might belong to a group not encountered in the training set. 

More formally, consider $\{(\mathbf{x}_1, \mathbf{l}_1),\ldots, (\mathbf{x}_N, \mathbf{l}_N)\}$ a complete set of learning observations, where $\mathbf{x}_n$ denotes a $p$-variate outcome and $\mathbf{l}_{n}$ its associated class label, such that $l_{ng}=1$ if observation $n$ belongs to group $g$ and $0$ otherwise, $g=1,\ldots, G$. Analogously, let $\{(\mathbf{y}_1, \mathbf{z}_1),\ldots,$ $(\mathbf{y}_M, \mathbf{z}_M)\}$ the set of unlabelled observations $\mathbf{y}_m$ with unknown classes $\mathbf{z}_{m}$, where  $z_{mg}=1$ if observation $m$ belongs to group $g$ and $0$ otherwise, $g=1,\ldots, E$, with $E \geq G$. Both $\mathbf{x}_n$, $n=1,\ldots,N$, and $\mathbf{y}_m$, $m=1,\ldots,M$, are assumed to be independent realizations of a continuous random vector $\mathcal{X} \in \mathbb{R}^p$; while $\mathbf{l}_{n}$ and $\mathbf{z}_{m}$ are considered to be realizations of a discrete random vector $\mathcal{C} \in \{1,\ldots, E \}$. Note that only $G$ classes, with $G$ possibly smaller than $E$, were encountered in the learning data. That is, there might be a number $H$ of ``hidden'' classes in the test such that $E=G + H$, with $H \geq 0$. Therefore, the marginal density for $\mathcal{X}$ is equal to:
\begin{equation*} \label{mixt_density}
f(\mathbf{x};\boldsymbol{\Theta})=\sum_{g=1}^E \tau_g f(\mathbf{x},\boldsymbol{\theta}_g),
\end{equation*}  
where $\tau_g$ is the prior probability of observing class $g$, such that $\sum_{g=1}^E\tau_g=1$, $f(\cdot,\boldsymbol{\theta}_g)$ is the density of the $g$th component of the mixture, parametrized by $\boldsymbol{\theta}_g$, and $\boldsymbol{\Theta}$ represents the collection of parameters to be estimated, $\boldsymbol{\Theta}= \{ \tau_1, \ldots,  \tau_E, \boldsymbol{\theta}_1, \ldots, \boldsymbol{\theta}_E \}$. 
Under the given framework, the \textit{observed log-likelihood} for the set of available information $\{(\mathbf{x}_n, \mathbf{l}_n, \mathbf{y}_m)\}$, $n=1,\ldots,N$, $m=1,\ldots, M$,  assumes the form:
\begin{align}\label{obs_ll}
\begin{split}
\ell(\boldsymbol{\Theta}| \mathbf{X}, \mathbf{Y},\mathbf{l})&=
\sum_{n=1}^N \sum_{g=1}^G l_{ng} \log{\left[\tau_g f(\mathbf{x}_n; \boldsymbol{\theta}_g)\right]} +\\
&+ \sum_{m=1}^M \log{\left[\sum_{g=1}^E\tau_g f(\mathbf{y}_m; \boldsymbol{\theta}_g)\right]}.
\end{split}
\end{align}
The first term in \eqref{obs_ll} accounts for the joint distribution of $(\mathbf{x}_n, \mathbf{l}_n)$, since both are observed; whereas in the second term only the marginal density of $\mathbf{y}_m$ contributes to the likelihood, given that its associated label $\mathbf{z}_m$ is unknown.
 Two alternative EM-based approaches for maximizing \eqref{obs_ll} with respect to $\boldsymbol{\Theta}$ in the case of Gaussian mixture are proposed in \cite{Bouveyron}. 
The adapted classifier assigns a new observation $\mathbf{y}_m$ to a known or previously unseen class with the associated highest posterior probability:
\begin{equation*} \label{MAP}
\hat{z}_{mg}=\mathbb{P}(\mathcal{C}=g|\mathcal{X}=\textbf{y}_m)=\frac{\hat{\tau}_g f(\mathbf{y}_m; \hat{\boldsymbol{\theta}}_g)}{\sum_{j=1}^E\hat{\tau}_j f(\mathbf{y}_m; \hat{\boldsymbol{\theta}}_j)},
\end{equation*}
for $g=1,\ldots, G,G+1,\ldots, E$. Note that the total number $E$ of groups is not established in advance and needs to be estimated: classical tools for model selection in the mixture model framework serve to this purpose \citep{A1974,Schwarz1978}.

The present paper extends the original AMDA model, briefly summarized in this Section, in three ways. Firstly, we account for both attribute and class noise that can appear in the samples \citep{Zhu}, employing impartial trimming \citep{Gordaliza1991}. Secondly, we consider a more flexible class of learners with the parsimonious parametrization based on the eigen-decomposition of \cite{Banfield1993} and \cite{Celeux1995}. Thirdly, we deal with a constrained parameter estimation 
to avoid convergence to degenerate solutions and to protect the estimates from spurious local maximizers that are likely to arise when searching for unobserved classes (see Section \ref{sec:eigen_ratio}).

The extended model is denoted as Robust and Adaptive Eigen Decomposition Discriminant Analysis (RAEDDA); its formulation, inferential aspects and selection criteria are covered in the next Section.

\section{Robust and Adaptive EDDA} \label{sec:RAEDDA}
\begin{figure*}
\centering
\vspace*{-2cm}
\hspace*{-.5cm} \includegraphics[scale=.9]{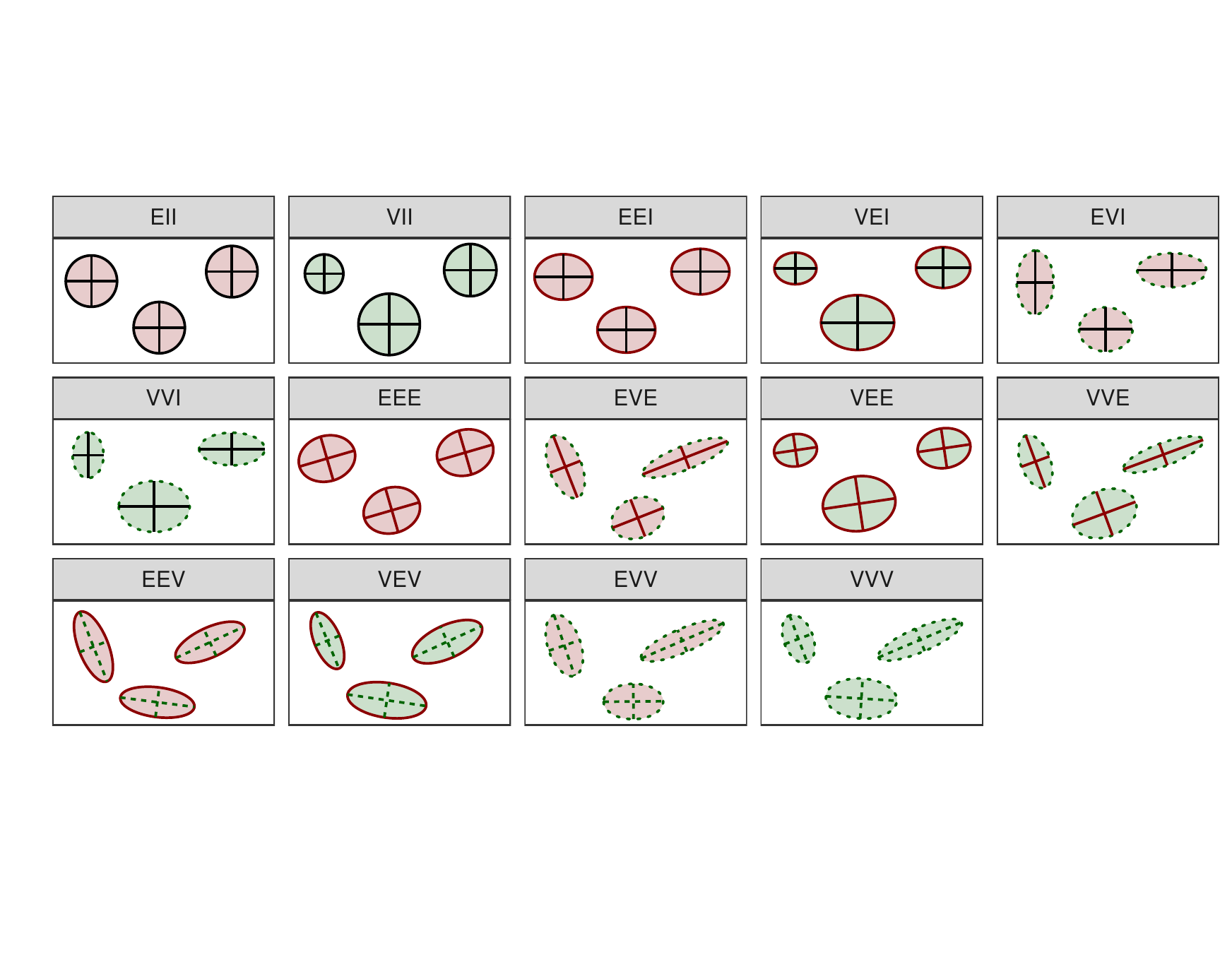}
\vspace*{-2.5cm}
\caption{Ellipses of isodensity for each of the 14 Gaussian models obtained by eigen-decomposition in case of three groups in two dimensions. Green (red) area denotes variable (equal) volume across components. Dashed green (solid red) perimeter denotes variable (equal) shape across components. Dashed green (solid red) axes denote variable (equal) orientation across components. Solid black perimeter denotes spherical shape. Solid black axes denote axis-aligned orientation.}
\label{fig:mclust_model}
\end{figure*}
\subsection{Model Formulation}
In this Section we introduce the new procedure, based on the definition of \textit{trimmed log-likelihood} \citep{Neykov2007} under a Gaussian mixture framework. 
Given a sample of $N$ training and $M$ test data, we construct a procedure for maximizing the \textit{trimmed observed data log-likelihood:}
\begin{eqnarray} \label{trim_ll}
\begin{split}
\ell_{trim}(&\boldsymbol{\tau}, \boldsymbol{\mu}, \boldsymbol{\Sigma}| \mathbf{X}, \mathbf{Y},\mathbf{l})=\\
&=\sum_{n=1}^N \zeta(\mathbf{x}_n)\sum_{g=1}^G \text{l}_{ng} \log{\left(\tau_g \phi(\mathbf{x}_n; \boldsymbol{\mu}_g, \boldsymbol{\Sigma}_g)\right)}+\\
& + \sum_{m=1}^M \varphi(\mathbf{y}_m)\log{\left(\sum_{g=1}^E\tau_g \phi(\mathbf{y}_m; \boldsymbol{\mu}_g, \boldsymbol{\Sigma}_g)\right)}
\end{split}
\end{eqnarray}
where $\phi(\cdot; \boldsymbol{\mu}_g, \boldsymbol{\Sigma}_g)$ represents the multivariate Gaussian density with mean vector $\boldsymbol{\mu}_g$ and covariance matrix $\boldsymbol{\Sigma}_g$; the functions \(\zeta(\cdot)\) and \(\varphi(\cdot)\) are indicator functions that determine whether each observation contributes or not to the trimmed likelihood, such that only \(\sum_{n=1}^N \zeta(\mathbf{x}_n)=\lceil N(1-\alpha_{l})\rceil\) and \(\sum_{m=1}^M \varphi(\mathbf{y}_m)= \lceil M(1-\alpha_{u})\rceil\) terms are not null in \eqref{trim_ll}. The \textit{labelled trimming level} \(\alpha_{l}\) and the \textit{unlabelled trimming level} \(\alpha_{u}\) identify the fixed fraction of observations, respectively belonging to the training and test sets, that are tentatively assumed to be unreliable at each iteration during the maximization of \eqref{trim_ll}. Once the trimming levels are specified, the proposed maximization process returns robustly estimated parameter values (see Sections \ref{sec:EM_transductive} and \ref{sec:EM_inductive} for details). 
Finally notice that only $G$ groups in \eqref{trim_ll}, out of the $E\geq G$ present in the population, are already captured within the labelled units, as in the AMDA model.

To introduce flexibility and parsimony, we consider the eigen-decomposition for the covariance matrices of \cite{Banfield1993} and \cite{Celeux1995}:
\begin{equation} \label{sigma_dec}
\boldsymbol{\Sigma}_g=\lambda_g\boldsymbol{D}_g\boldsymbol{A}_g\boldsymbol{D}^{'}_g
\end{equation}
where $\lambda_g=|\boldsymbol{\Sigma}_g|^{1/p}$, with $|\cdot|$ denoting the determinant, $\boldsymbol{A}_g$ is the scaled ($|\boldsymbol{A}_g|=1$) diagonal matrix of eigenvalues of $\boldsymbol{\Sigma}_g$ sorted in decreasing order and
$\boldsymbol{D}_g$ is a $p \times p$ orthogonal matrix whose columns are the normalized eigenvectors of $\boldsymbol{\Sigma}_g$, ordered according to their eigenvalues \citep{Greselin2013}. These elements correspond respectively to the orientation, shape and volume (alternatively called scale) of the Gaussian components. By imposing cross-constraints on the parameters in \eqref{sigma_dec} 14 patterned models can be defined: their nomenclature and  characteristics are 
 represented in Figure \ref{fig:mclust_model}.
\cite{Bensmail1996} defined a family of supervised classifiers based on such decomposition, known in the literature as Eigenvalue Decomposition Discriminant Analysis (EDDA). Our proposal generalizes the original EDDA including robust estimation and adaptive learning, hence the name Robust and Adaptive Eigenvalue Decomposition Discriminant Analysis (RAEDDA). 
Two alternative estimation procedures for maximizing  \eqref{trim_ll} are proposed. The transductive approach (see Section \ref{sec:EM_transductive}) works on the simultaneous usage of learning and test sets to estimate model parameters. The inductive approach (see Section \ref{sec:EM_inductive}), instead, consists of two distinctive phases: in the first one the training set is employed for estimating parameters of the $G$ known groups; 
in the second phase the unlabelled observations are assigned to the known groups
whilst searching for new classes and estimating their parameters. 
Computational aspects for both procedures are detailed in the next Sections.

\subsection{Estimation Procedure: Transductive Approach} \label{sec:EM_transductive}
\begin{figure}
\centering
\includegraphics[scale=0.8]{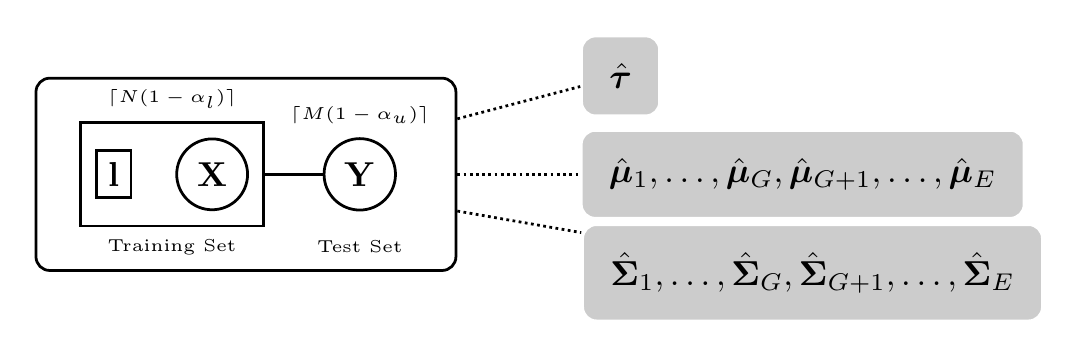}
\caption{General framework of the \textit{robust transductive estimation} approach: $\lceil N(1-\alpha_{l})\rceil$ observations in the training and $\lceil M(1-\alpha_{u})\rceil$ observations in the test are jointly employed in estimating parameters for the known and hidden classes.}
\label{fig:transductive_approach}
\end{figure}

Transductive inference considers the joint exploitation of training and test sets to solve a specific learning problem \citep{Vapnik2000, Kasabov2003}. Transductive reasoning is applied for instance in semi-supervised classification methods: the data generating process is assumed to be the same for labelled and unlabelled observations and hence units coming from both sets can be used to build the classification rule. Within the family of model-based classifiers, examples of such methods are the updating classification rules by \cite{Dean2006} and its robust generalization introduced in \cite{Cappozzo2019b}. The present context is more general than semi-supervised learning since the total number of classes $E$ might be strictly larger than the $G$ ones observed in the training set. Therefore, an ad-hoc procedure needs to be introduced: a graphical representation of the transductive approach is reported in Figure \ref{fig:transductive_approach}.

An adaptation of the EM algorithm \citep{Dempster1977} that includes a Concentration Step \citep{Driessen1999} and an eigenvalue-ratio restriction \citep{Ingrassia2004} is employed for maximizing \eqref{trim_ll}. The former serves the purpose of enforcing impartial trimming in both labelled and unlabelled units at each step of the algorithm, whereas the latter prevents the procedure to be trapped in spurious local maximizers that may arise whenever a random pattern in the test is wrongly fitted to form a hidden class (see Section \ref{sec:eigen_ratio}). Particularly, the considered eigenvalue-ratio restriction is as follows:
\begin{equation} \label{eigen_contr}
\Pi/\pi\leq c
\end{equation}
where 
\[\Pi=\max_{g=1\ldots E}\max_{l=1\ldots p} d_{lg}\] and 
\[\pi=\min_{g=1\ldots E}\min_{l=1\ldots p} d_{lg},\]
with \(d_{lg}\), $l=1,\ldots, p$ being the eigenvalues of the matrix
\(\boldsymbol{\Sigma}_g\) and $c \geq1$ being a fixed constant	 \citep{Ingrassia2004}. Constraint \eqref{eigen_contr} simultaneously controls differences between groups and departures from sphericity, by forcing the relative length of the axes of the equidensity ellipsoids of the multivariate normal distribution (modeling each group) to be smaller than $\sqrt{c}$ \citep{Garcia-Escudero2014}. From the seminal paper of \cite{day1969estimating} we know that the likelihood for a Gaussian mixture is unbounded and the ML approach is thus an ill posed problem, whenever constraints like in \eqref{eigen_contr} are not assumed.
Notice further that the constraint in \eqref{eigen_contr} is still needed whenever either the shape or the volume is free to vary across components \citep{Garcia-Escudero}: feasible algorithms for enforcing the eigen-ratio constraint under different specification of the covariance matrices as per Figure \ref{fig:mclust_model} have been derived in \cite{Cappozzo2019b}.

Under the transductive learning phase, the \textit{trimmed complete data log-likelihood} reads as:

\begin{eqnarray} \label{trim_comp_ll}
\begin{split}
\ell_{trim_c}&(\boldsymbol{\tau}, \boldsymbol{\mu}, \boldsymbol{\Sigma}| \mathbf{X}, \mathbf{Y},\mathbf{l}, \mathbf{z})=\\
&=\sum_{n=1}^N \zeta(\mathbf{x}_n)\sum_{g=1}^G \text{l}_{ng} \log{\left(\tau_g \phi(\mathbf{x}_n; \boldsymbol{\mu}_g, \boldsymbol{\Sigma}_g)\right)}+\\
& + \sum_{m=1}^M \varphi(\mathbf{y}_m) \sum_{g=1}^E z_{mg}\log{\left(\tau_g \phi(\mathbf{y}_m; \boldsymbol{\mu}_g, \boldsymbol{\Sigma}_g)\right)}
\end{split}
\end{eqnarray}
The following steps detail a constrained EM algorithm for jointly estimating model parameters (see Figure \ref{fig:transductive_approach}) whilst searching for new classes and outliers.

Unlike what is suggested in \cite{Bouveyron}, the EM initialization is here performed in two subsequent steps for preventing outliers to spoil the starting values and henceforth driving the entire algorithm to reach uninteresting solutions. We firstly make use of a robust procedure to obtain a set of parameter estimates $\{\bar{\boldsymbol{\tau}},\bar{\boldsymbol{\mu}},\bar{\boldsymbol{\Sigma}}\}$ for the known groups $G$ using only the training set. Afterwards, if $E>G$, we randomly initialize the parameters for the $H=E-G$ hidden classes taking advantage of the known groups structure learned in the previous step. Notice that, as in \cite{Bouveyron}, at this moment the number of hidden classes $E$ is assumed to be known: we will discuss its estimation later (see Section \ref{sec:TBIC}).
\begin{itemize}
\item
\emph{Robust Initialization for the $G$ known groups}: set \(k=0\). Employing only the labelled data, we obtain robust starting values for $\boldsymbol{\mu}_g$ and $\boldsymbol{\Sigma}_g$, $g=1,\ldots,G$ as follows:
  \begin{enumerate}
\item For each known class $g$, draw a random $(p + 1)$-subset $J_g$ and compute its empirical mean $\bar{\boldsymbol{\mu}}^{(0)}_g
$ and covariance $\bar{\boldsymbol{\Sigma}}^{(0)}_g$ according to the considered parsimonious structure.
\item Set
\begin{align*}
\begin{split}
\{\bar{\boldsymbol{\tau}},\bar{\boldsymbol{\mu}},\bar{\boldsymbol{\Sigma}}\}=\{ \bar{\tau}_1,\ldots,\bar{\tau}_G, \bar{\boldsymbol{\mu}}_1,\ldots,\bar{\boldsymbol{\mu}}_G,
\bar{\boldsymbol{\Sigma}}_1,\ldots,\bar{\boldsymbol{\Sigma}}_G\}=\\
=\{\bar{\tau}_1^{(0)},\ldots,\bar{\tau}_G^{(0)}, \bar{\boldsymbol{\mu}}_1^{(0)},\ldots,\bar{\boldsymbol{\mu}}_G^{(0)},
\bar{\boldsymbol{\Sigma}}^{(0)}_1,\ldots,\bar{\boldsymbol{\Sigma}}_G^{(0)} \}
\end{split}
\end{align*}
where $\bar{\tau}_1^{(0)}=\ldots=\bar{\tau}_G^{(0)}=1/G$.
\item 
For each $\mathbf{x}_n$, $n=1,\ldots,N$, compute the conditional density
\begin{equation}\label{cond_dens}
f(\mathbf{x}_n|l_{ng}=1; \bar{\boldsymbol{\mu}},\bar{\boldsymbol{\Sigma}})=\phi\left(\mathbf{x}_n; \bar{\boldsymbol{\mu}}_g, \bar{\boldsymbol{\Sigma}}_g \right) \:\:\:\:\: g=1,\ldots, G.
\end{equation}
\(\lfloor N\alpha_{l} \rfloor \%\) of the samples with lower values of \eqref{cond_dens} are temporarily discarded from contributing to the parameters estimation. The rationale being that observations suffering from either class or attribute noise are unplausible under the currently fitted model. That is, $\zeta(\mathbf{x}_n)=0$ in \eqref{trim_comp_ll} for such observations.
\item The parameter estimates for the $G$ known classes are updated,
  based on the non-discarded observations:
  \begin{align}
  \bar{\tau}_g=\frac{\sum_{n=1}^N \zeta(\mathbf{x}_n)l_{ng}}{\lceil N(1-\alpha_{l})\rceil}\:\:\:\:\: g=1,\ldots, G\\
  \bar{\boldsymbol{\mu}}_g=\frac{\sum_{n=1}^N \zeta(\mathbf{x}_n)l_{ng}\mathbf{x}_n}{\sum_{n=1}^N\zeta(\mathbf{x}_n)l_{ng}}\:\:\:\:\: g=1,\ldots, G.
    \end{align}
  Estimation of $\boldsymbol{\Sigma}_g$ depends on the considered patterned model, details are given in \cite{Bensmail1996}.
  \item Iterate $3-4$ until the $\lfloor N\alpha_{l} \rfloor$ discarded observations are exactly the same on two consecutive iterations, then stop.
\end{enumerate}
The procedure described in steps $1-5$ is performed $\texttt{n\_init}$ times, and the parameter estimates $\{\bar{\boldsymbol{\tau}},\bar{\boldsymbol{\mu}},\bar{\boldsymbol{\Sigma}}\}$ that lead to the highest value of the objective function 
\[\ell_{trim}(\bar{\boldsymbol{\tau}}, \bar{\boldsymbol{\mu}}, \bar{\boldsymbol{\Sigma}}| \mathbf{X}, \mathbf{\text{l}})= \sum_{n=1}^N \zeta(\mathbf{x}_n)\sum_{g=1}^G\text{l}_{ng} \log{\left[\bar{\tau}_g \phi(\mathbf{x}_n; \bar{\boldsymbol{\mu}}_g, \bar{\boldsymbol{\Sigma}}_g)\right]}\] are retained. Therefore, $\{\bar{\boldsymbol{\tau}},\bar{\boldsymbol{\mu}},\bar{\boldsymbol{\Sigma}}\}$ is the output of the robust initialization phase for the $G$ known classes. 

\item
\emph{Initialization for the $H$ hidden classes}: If $E>G$, starting values for the $H=E-G$ hidden classes need to be properly initialized as follows:
\begin{enumerate}
\item For each hidden class $h$, $h=G+1, \ldots,E$, draw a random $(p + 1)$-subset $J_h$ and compute its empirical mean $\hat{\boldsymbol{\mu}}^{(0)}_h
$ and variance covariance matrix $\hat{\boldsymbol{\Sigma}}^{(0)}_h$ according to the considered parsimonious structure. Mixing proportions $\tau_h$ are drawn from $\mathcal{U}_{[0,1]}$ and initial values set equal to
\[\hat{\tau}^{(0)}_h=\frac{\tau_h}{\sum_{j=G+1}^E\tau_j}\times\frac{H}{E}, \:\:h=G+1,\ldots,E.\]
The previously estimated $\tau_g$ should also be renormalized:
\[\hat{\tau}^{(0)}_g =\bar{\tau}_g\times\frac{G}{E}, \:\:g=1,\ldots, G,\]
to obtain that the initialized vector of mixing proportion sums to $1$ over the $E$ groups.   
\end{enumerate}
\item
If the selected patterned model allows for heteroscedastic $\boldsymbol{\Sigma}_g$, and  $\hat{\boldsymbol{\Sigma}}_g^{(0)}$, $g=1,\ldots,E$ do not satisfy \eqref{eigen_contr}, constrained maximization needs to be enforced. Given the semi-supervised nature of the problem at hand, we propose to further rely on the information that can be extracted from the robustly initialized estimates $\{\bar{\boldsymbol{\tau}},\bar{\boldsymbol{\mu}},\bar{\boldsymbol{\Sigma}}\}$ to set sensible values for the fixed constant  $c \geq1$ required in the eigenvalue-ratio restriction. That is, if no prior information for the value $c$ is available, as it is almost always the case in real applications \citep{Garcia-Escudero}, the following quantity could be, al least initially, used:
\begin{equation} \label{c_tilde}
\tilde{c}=\dfrac{\max_{g=1\ldots G}\max_{l=1\ldots p} \bar{d}_{lg}}{\min_{g=1\ldots G}\min_{l=1\ldots p} \bar{d}_{lg}}
\end{equation}
with \(\bar{d}_{lg}\), $l=1,\ldots, p$ being the eigenvalues of the matrix
\(\bar{\boldsymbol{\Sigma}}_g\), $g=1,\ldots,G$. 
This implicitly means that we expect extra hidden groups whose difference among group scatters is no larger than that observed for the known groups. Such a choice prevents the appearance of spurious solutions, protecting the adapted learner to wrongly identify random patterns as unobserved classes whilst allowing for groups variability to be preserved. Nevertheless, one might want to allow more flexibility in the group structure and use \eqref{c_tilde} as a lower bound for $c$, rather than an actual reasonable value. Once having obtained $\hat{\boldsymbol{\Sigma}}_g^{(0)}$ under the eigenvalue ratio constraint, the following EM iterations produce an algorithm that maximizes the observed likelihood in \eqref{trim_ll}.

\item
   \emph{EM Iterations:} denote by \[\hat{\boldsymbol{\Theta}}^{(k)}=\{ \hat{\tau}_1^{(k)},\ldots,\hat{\tau}_E^{(k)}, \hat{\boldsymbol{\mu}}_1^{(k)},\ldots,\hat{\boldsymbol{\mu}}_E^{(k)},
\hat{\boldsymbol{\Sigma}}^{(k)}_1,\ldots,\hat{\boldsymbol{\Sigma}}_E^{(k)} \} \]  the parameter estimates at the $k$-th iteration of the algorithm.

\begin{itemize}
\item
  \emph{Step 1 - Concentration}: the trimming procedure is implemented by
  discarding the \(\lfloor N\alpha_{l} \rfloor\) observations \(\mathbf{x}_n\) with smaller values of
  \begin{equation} \label{trim_l}
  D\left(\mathbf{x}_n; \hat{\boldsymbol{\Theta}}^{(k)} \right)=\prod_{g=1}^E \left[ \phi \left(\mathbf{x}_n; \hat{\boldsymbol{\mu}}^{(k)}_g, \hat{\boldsymbol{\Sigma}}^{(k)}_g \right)\right]^{l_{ng}} \:\:\:\:\: n=1,\ldots,N
    \end{equation}
  and discarding the \(\lfloor M\alpha_{u}\rfloor\) observations \(\mathbf{y}_m\)  with smaller values of
  \begin{equation} \label{trim_u}
D\left(\mathbf{y}_m; \hat{\boldsymbol{\Theta}}^{(k)}\right)=\sum_{g=1}^E \hat{\tau}^{(k)}_g \phi \left(\mathbf{y}_m; \hat{\boldsymbol{\mu}}^{(k)}_g, \hat{\boldsymbol{\Sigma}}^{(k)}_g \right) \:\:\:\:\: m=1,\ldots,M.
 \end{equation}
Namely, we set $\zeta(\cdot)=0$ and $\varphi(\cdot)=0$ in \eqref{trim_comp_ll} for the trimmed units in the training and test sets, respectively. Notice that we implicitly impose $l_{ng}=0$ $\forall \:n=1,\ldots,N$, $g=G+1,\ldots,E$ in \eqref{trim_l}. That is, none of the learning units belong to one of the hidden classes $h$, $h=G+1,\ldots, E$.
\item
  \emph{Step 2 - Expectation}: for each non-trimmed observation \(\mathbf{y}_m\)
  compute the posterior probabilities
  \begin{equation} \label{trans_MAP}
\hat{z}_{mg}^{(k+1)}=\frac{\hat{\tau}^{(k)}_g \phi \left(\mathbf{y}_m; \hat{\boldsymbol{\mu}}^{(k)}_g, \hat{\boldsymbol{\Sigma}}^{(k)}_g \right)}{D\left(\mathbf{y}_m; \hat{\boldsymbol{\theta}}^{(k)}\right)} 
\end{equation}
for $g=1,\ldots, E$, $m=1,\ldots, M$.
\item
  \emph{Step 3 - Constrained Maximization}: the parameter estimates are updated,
  based on the non-discarded observations and the current estimates for the unknown labels:
  \begin{equation*}
  \hat{\tau}_g^{(k+1)}=\frac{\sum_{n=1}^N \zeta(\mathbf{x}_n)l_{ng}+ \sum_{m=1}^M \varphi(\mathbf{y}_m)\hat{z}_{mg}^{(k+1)}}{\lceil N(1-\alpha_{l})\rceil+\lceil M(1-\alpha_{u})\rceil}  
  \end{equation*}
  \begin{equation*}
  \hat{\boldsymbol{\mu}}_g^{(k+1)}=\frac{\sum_{n=1}^N \zeta(\mathbf{x}_n)l_{ng}\mathbf{x}_n+\sum_{m=1}^M\varphi(\mathbf{y}_m)\hat{z}_{mg}^{(k+1)}\mathbf{y}_m}{\sum_{n=1}^N\zeta(\mathbf{x}_n)l_{ng}+\sum_{m=1}^M\varphi(\mathbf{y}_m)\hat{z}_{mg}^{(k+1)}}  
  \end{equation*}
for $g=1,\ldots, E$. Estimation of $\boldsymbol{\Sigma}_g$ depends on the considered patterned model and on the eigenvalues ratio constraint. Details are given in \cite{Bensmail1996} and, if \eqref{eigen_contr} is not satisfied, in Appendix C of  \cite{Cappozzo2019b}.
\item

  \emph{Step 4 - Convergence of the EM algorithm}: if convergence has not been reached (see Section \ref{sec:convergence}), set \(k=k+1\) and repeat steps 1-4.
 \end{itemize}  
\end{itemize}
Notice that, once the hidden classes have been properly initialized, the transductive approach relies on an EM algorithm that makes use of both training and test sets for jointly estimating the parameters of known and hidden classes, with no distinction between the two. The final output from the procedure is a set of parameters $\{\hat{\tau}_g,\hat{\boldsymbol{\mu}}_g,\hat{\boldsymbol{\Sigma}}_g\}$, $g=1,\ldots,E$, and values for the indicator functions \(\zeta(\cdot)\) and \(\varphi(\cdot)\). 
Furthermore, the estimated values \(\hat{z}_{mg}\) provide a classification for the unlabelled observations \(\mathbf{y}_m\) using the MAP rule.

Summing up, the procedure identifies a mislabelled and/or an outlying unit in the training set when \(\zeta(\mathbf{x}_n)=0\), an outlier in the test set  when \(\varphi(\mathbf{y}_m)=0\) and an observation in the test belonging to a hidden class whenever \(\text{argmax}_{g=1,\ldots,E}\hat{z}_{mg} \in \{G+1,\ldots,E\}$. As appropriately pointed out by an anonymous reviewer, one may be interested in distinguishing between training units that were trimmed due to their attribute and/or class noise; thus possibly reassigning correct labels to the latter subgroup. Clearly, \eqref{trans_MAP} may be used for this purpose, even though we argue that extra care needs to be applied to avoid that outlying observations are also classified according to the MAP rule. A simple two-step proposal to limit such risk and to identify wrongly labeled units is as follows: firstly every trimmed training unit is a-posteriori reassigned using \eqref{trans_MAP} and, secondly, \eqref{trim_l} is evaluated employing the final EM estimates and the newly reassigned class label. At this point, only the observations displaying higher value than the $\alpha_l$-quantile considered in the trimming step
will be assigned to the estimated classes; along the lines of what done in the discovery phase of the inductive approach (see Section \ref{sec:disc_phase}). As a last worthy comment we remark that, from a practitioner perspective, it may be relevant to reassign a trimmed unit to its correct class only after careful study has been devoted to the analysis of the discarded subset: unraveling the cause of the mislabeling process could be of great importance and, unfortunately, no algorithm can automatically unmask that.
\subsection{Estimation Procedure: Inductive Approach} \label{sec:EM_inductive}
\begin{figure*}
\centering
\hspace*{-.5cm}
\includegraphics[scale=0.9]{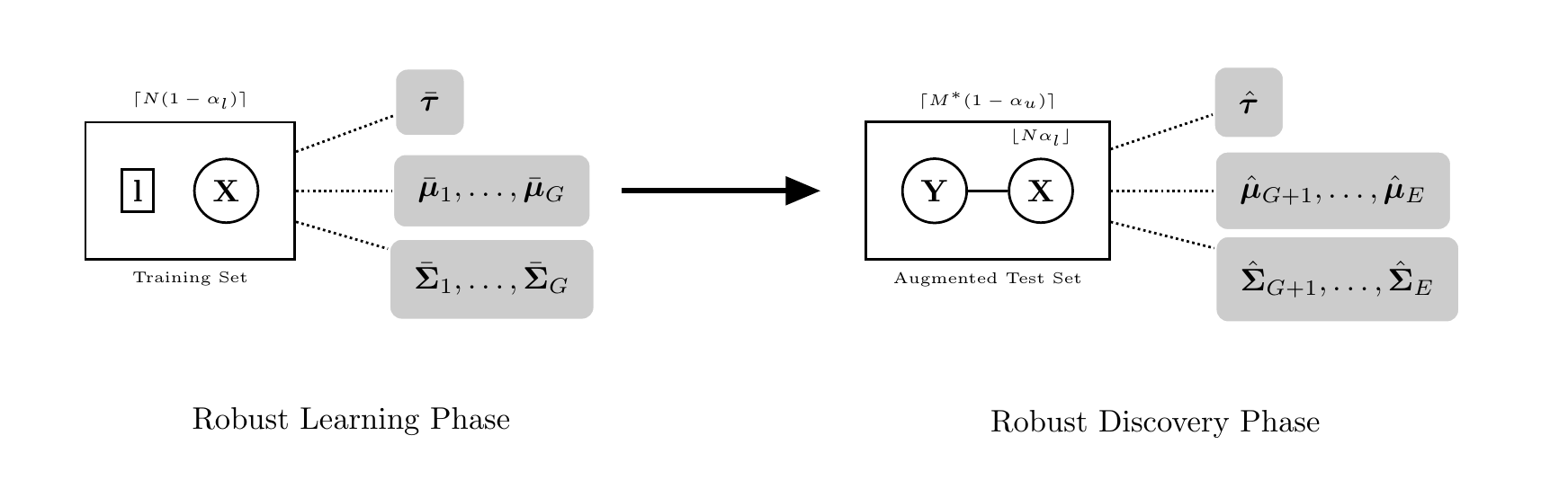}
\caption{General framework of the \textit{robust inductive estimation} approach. $\lceil N(1-\alpha_{l})\rceil$ observations in the training are used to estimate parameters for the known groups in the Robust Learning Phase. Keeping fixed the estimates obtained in the previous phase, $\lceil M^*\left(1- \alpha_u \right) \rceil$ observations in the augmented test are then employed in estimating parameters only for the hidden classes, $M^*=M+\lfloor N\alpha_l\rfloor$.}
\label{fig:inductive_approach}
\end{figure*}
In contrast with transductive inference, the inductive learning approach aims at solving a broader type of problem: a general model is built from the training set to be ideally applied on any new data point, without the need of a specific test set to be previously defined \citep{Mitchell:1997:ML:541177, Pang2004}. As a consequence, this approach is most suitable for real-time dynamic classification of data streams, since only the classification rule (i.e., model parameters) is stored and the training set need not be kept in memory. Operationally, inductive learning is performed in two steps: a robust learning phase and a robust discovery phase (see Figure \ref{fig:inductive_approach}). In the learning phase, only training observations are considered and we therefore fall into the robust fully-supervised framework for classification. In the robust discovery phase only the parameters for the $E-G$ extra
classes need to be estimated, since the parameters obtained in the learning phase are kept fixed. The entire procedure is detailed in the next Sections. 
\subsubsection{Robust learning phase} \label{sec:EM_inductive_l}
The first phase of the inductive approach consists of estimating parameters for the observed classes using only the training set. That is, we aim at building a robust fully-supervised model considering only the (complete set) of observations $\{\mathbf{x}_n,\mathbf{l}_n\}, n=1,\ldots,N$. The associated \textit{trimmed log-likelihood} to be maximized with respect to parameters $\{\tau_g,\boldsymbol{\mu}_g,\boldsymbol{\Sigma}_g\}$, $g=1,\ldots,G$, reads:
\begin{align}
\begin{split}
\label{trim_ll_learning}
\ell_{trim}(&\boldsymbol{\tau}, \boldsymbol{\mu}, \boldsymbol{\Sigma}| \mathbf{X}, \mathbf{\text{l}})=\\
=&\sum_{n=1}^N \zeta(\mathbf{x}_n)\sum_{g=1}^G \text{l}_{ng} \log{\left(\tau_g \phi(\mathbf{x}_n; \boldsymbol{\mu}_g, \boldsymbol{\Sigma}_g)\right)}
\end{split}
\end{align}
Notice that \eqref{trim_ll_learning} is the first of the two addends that compose \eqref{trim_ll}. In this situation, the obtained model is the Robust Eigenvalue Decomposition Discriminant Analysis \citep{Cappozzo2019b}. The estimation procedure coincides with the \emph{Robust Initialization for the $G$ known groups} step in the transductive approach (see Section \ref{sec:EM_transductive}). 

At this point, one could employ the trimmed adaptation of the Bayesian Information Criterion \citep{Neykov2007, Schwarz1978, Fraley2002} for selecting the best model among the 14 covariance decompositions of Figure \ref{fig:mclust_model}. Notice that the parametrization chosen in the learning phase will influence the available models for the discovery phase (see Figure \ref{fig:partial_order_mclust}). 

This concludes the learning phase and the role the training set has in the estimation procedure: from now on $\{\mathbf{x}_n,\mathbf{l}_n\}, n=1,\ldots,N$ may be discarded. The only exception being the $\lfloor N \alpha_l \rfloor$ units for which $\zeta(\mathbf{x}_n)=0$: denote such observations with $\{\mathbf{x}^{*}_i,\mathbf{l}^{*}_i\}$, $i=1,\ldots,\lfloor N \alpha_l \rfloor$. These are the observations not included in the estimation procedure, that is, samples whose conditional density \eqref{cond_dens} is the smallest. This could be due to either a wrong label $\mathbf{l}^{*}_i$ or $\mathbf{x}_i^{*}$ to be an actual outlier: in the former case, $\mathbf{x}_i^{*}$ could still be potentially useful for detecting unobserved classes. We therefore propose to join the $\lfloor N \alpha_l \rfloor$ units excluded from the learning phase with the test set to define an \emph{augmented test set} $\mathbf{Y}^{*}=\mathbf{Y} \cup \mathbf{X}^{(\alpha_l)}$, with elements $\mathbf{y}^{*}_m$, $m=1,\ldots, M^{*}$, $M^{*}=\left( M+\lfloor N \alpha_l \rfloor \right)$, to be employed in the discovery phase. Clearly, $\mathbf{Y}^{*}$ reduces to $\mathbf{Y}$ if $\alpha_l=0$. In addition, depending on the real problem at hand, the recovery of the $\mathbf{x}_i^{*}$ units may be too time consuming or too costly or simply impossible when an online classification is to be performed. In such cases the robust discovery phase described in the next Section can still be applied making use of the original test units $\mathbf{y}_m$, $m=1,\ldots, M$.

\subsubsection{Robust discovery phase} \label{sec:disc_phase}
In the robust discovery phase, we search for $H=E-G$ hidden classes robustly estimating the related parameters in an unsupervised way. Particularly, the set $\{\bar{\boldsymbol{\mu}}_g,\bar{\boldsymbol{\Sigma}}_g\}$, $g=1,\ldots,G$ will remain fixed, as indicated by the bar in the notation, throughout the discovery phase. Therefore, the observed \textit{trimmed log-likelihood}, here given by

\begin{eqnarray} 
\begin{split} \label{trim_ll_discovery}
\ell_{trim}&(\boldsymbol{\tau}, \boldsymbol{\mu}, \boldsymbol{\Sigma}| \mathbf{Y}^{*}, \bar{\boldsymbol{\mu}}, \bar{\boldsymbol{\Sigma}})=\\
&=\sum_{m=1}^{M^{*}} \varphi(\mathbf{y}^{*}_m)\log \bigg( \sum_{g=1}^G\tau_g \phi(\mathbf{y}^{*}_m; \bar{\boldsymbol{\mu}}_g, \bar{\boldsymbol{\Sigma}}_g)+\\
& + \sum_{h=G+1}^E \tau_h \phi(\mathbf{y}^{*}_m; \boldsymbol{\mu}_h, \boldsymbol{\Sigma}_h) \bigg)
\end{split}
\end{eqnarray}
will be maximized with respect to $\{\boldsymbol{\tau},\{\boldsymbol{\mu}_h,\boldsymbol{\Sigma}_h\}_{h=G+1,\ldots,E}\}$. Direct maximization of \eqref{trim_ll_discovery} is an intractable problem, and we extend \cite{Bouveyron} making again use of an EM algorithm defining a proper \textit{complete trimmed log-likelihood}:
\begin{eqnarray} 
\begin{split} \label{trim_ll_discovery_complete}
\ell_{trim_{c}}(&\boldsymbol{\tau}, \boldsymbol{\mu}, \boldsymbol{\Sigma}| \mathbf{Y}^{*}, \bar{\boldsymbol{\mu}}, \bar{\boldsymbol{\Sigma}}, \mathbf{z}^{*})=\\
=&\sum_{m=1}^{M^{*}}  \varphi(\mathbf{y}^{*}_m)\bigg( \sum_{g=1}^G z_{mg}^{*} \log (\tau_g \phi(\mathbf{y}^{*}_m; \bar{\boldsymbol{\mu}}_g, \bar{\boldsymbol{\Sigma}}_g)) +\\
& + \sum_{h=G+1}^E z_{mh}^{*} \log ( \tau_h \phi(\mathbf{y}^{*}_m; \boldsymbol{\mu}_h, \boldsymbol{\Sigma}_h)) \bigg)
\end{split}
\end{eqnarray}
The following steps delineate the procedure needed for maximizing \eqref{trim_ll_discovery}:
\begin{itemize}
\item
\emph{Initialization for the $H$ hidden classes}: 
\begin{enumerate}
\item For each hidden class $h$, $h=G+1, \ldots,E$, draw a random $(p + 1)$-subset $J_h$ and compute its empirical mean $\hat{\boldsymbol{\mu}}^{(0)}_h
$ and variance covariance matrix $\hat{\boldsymbol{\Sigma}}^{(0)}_h$ according to the considered parsimonious structure. Mixing proportions $\tau_h$ are drawn from $\mathcal{U}_{[0,1]}$ and initial values set equal to
\[\hat{\tau}^{(0)}_h=\frac{\tau_h}{\sum_{j=G+1}^E\tau_j}\times\frac{H}{E}, \:\:h=G+1,\ldots,E.\]
 The $\tau_g$ estimated in the robust learning phase should also be renormalized:
 \[\hat{\tau}^{(0)}_g =\bar{\tau}_g \times \frac{G}{E}, \:\:g=1,\ldots, G.\] 
\end{enumerate}

\item
If the selected patterned model allows for heteroscedastic $\boldsymbol{\Sigma}_g$, and  $\hat{\boldsymbol{\Sigma}}_g^{(0)}$, $g=G+1,\ldots,E$ do not satisfy \eqref{eigen_contr}, constrained maximization needs to be enforced. Notice that, thanks to the inductive approach, only the estimates for the $H$ hidden groups covariance matrices need to satisfy the eigen-ratio constraint. Moreover, the information extracted from the robust learning phase provides a lower bound, using \eqref{c_tilde}, for the fixed constant  $c \geq1$ required in the eigenvalue-ratio restriction. The implicit assumption that the hidden groups variability is no larger than that estimated for the known classes protects new estimates from spurious solutions: given the unsupervised nature of the problem they can easily arise when searching for unobserved classes also in the simplest scenarios (see Section \ref{sec:eigen_ratio}).
%

Once initial values have been determined for the parameters of the hidden classes, the following EM iterations maximize \eqref{trim_ll_discovery}.
\item
   \emph{EM Iterations:} denote by \[\hat{\boldsymbol{\Theta}}^{(k)}=\{ \hat{\tau}_1^{(k)},\ldots,\hat{\tau}_E^{(k)}, \hat{\boldsymbol{\mu}}_{G+1}^{(k)},\ldots,\hat{\boldsymbol{\mu}}_E^{(k)},
\hat{\boldsymbol{\Sigma}}^{(k)}_{G+1},\ldots,\hat{\boldsymbol{\Sigma}}_E^{(k)} \} \]  the parameter estimates at the $k$-th iteration of the algorithm.

\begin{itemize}
\item
  \emph{Step 1 - Concentration}: Define 
  \begin{equation*} 
 D_g\left( \mathbf{y}^*_m; \hat{\boldsymbol{\Theta}}^{(k)} \right)=
\begin{cases}
  \hat{\tau}^{(k)}_g \phi\left(\mathbf{y}^*_m; \bar{\boldsymbol{\mu}}_g, \bar{\boldsymbol{\Sigma}}_g \right) & g=1,\ldots, G \\
      \hat{\tau}^{(k)}_g \phi\left(\mathbf{y}^*_m; \hat{\boldsymbol{\mu}}^{(k)}_g, \hat{\boldsymbol{\Sigma}}^{(k)}_g \right) & g=G+1,\ldots, E \\
 \end{cases}
\end{equation*}
The trimming procedure is implemented by
  discarding the \(\lfloor M^{*} \alpha_{u}\rfloor\) observations \(\mathbf{y}^{*}_m\)  with smaller values of
\begin{equation*}   
D\left(\mathbf{y}_m^{*}; \hat{\boldsymbol{\Theta}}^{(k)}\right)= \sum_{g=1}^E D_g\left( \mathbf{y}^*_m; \hat{\boldsymbol{\Theta}}^{(k)} \right) \:\:\:\:\: m=1,\ldots,M^{*}.
\end{equation*}
That is, we set $\varphi(\cdot)=0$ in \eqref{trim_ll_discovery_complete} for the trimmed units in the augmented test set.
\item
  \emph{Step 2 - Expectation}: for each non-trimmed observation \(\mathbf{y}^*_m\)
  compute the posterior probabilities
  \begin{equation*}
\hat{z}_{mg}^{*^{(k+1)}}=\frac{D_g\left( \mathbf{y}^*_m; \hat{\boldsymbol{\Theta}}^{(k)} \right)}{D\left(\mathbf{y}^*_m; \hat{\boldsymbol{\Theta}}^{(k)}\right)} \:\:\:\:\: g=1,\ldots, E; \:\:\:\: m=1,\ldots, M^*.
\end{equation*}
\begin{figure}
\includegraphics[scale=.5]{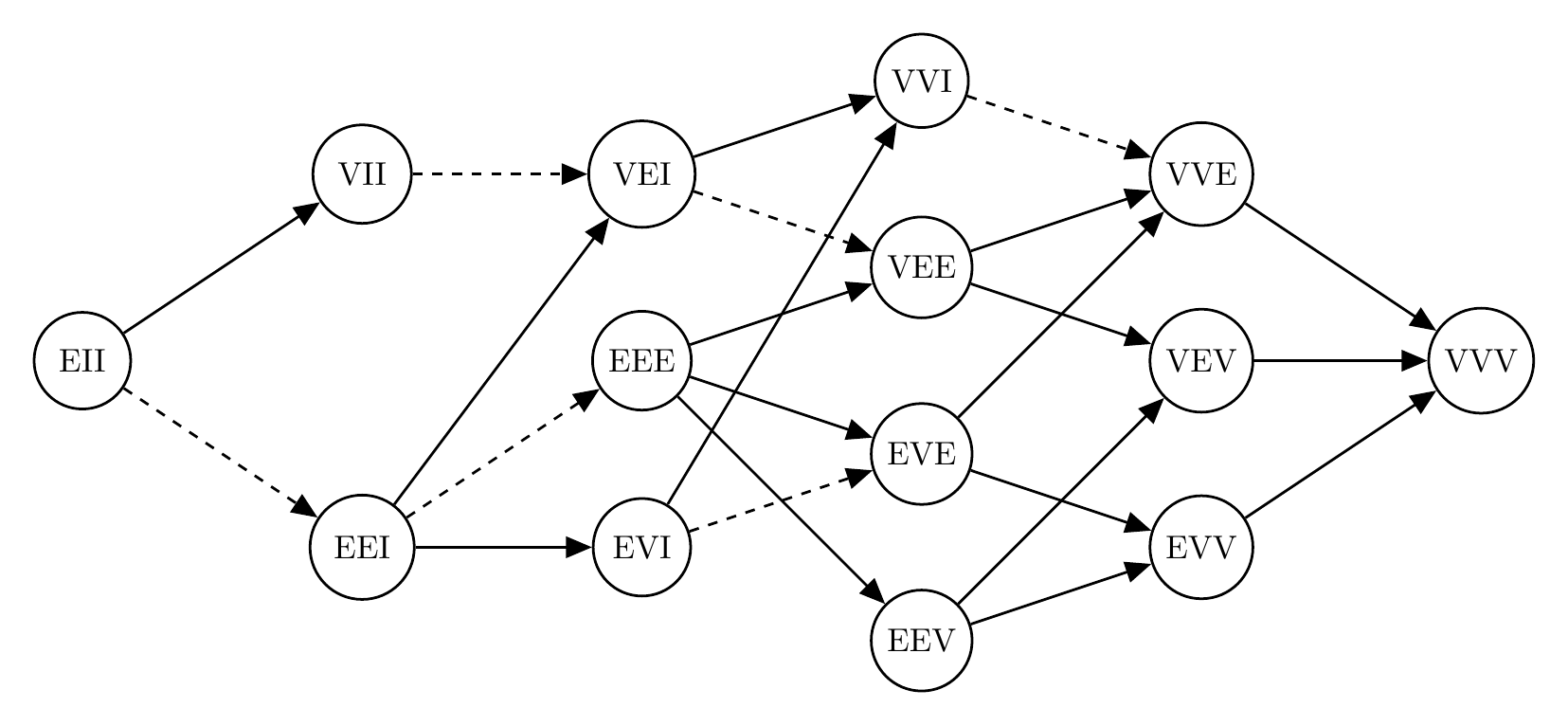}
\centering
\caption{Partial-order structure in the eigen-decomposition for the covariance matrices. Model complexity increases from left to right. Dashed arrows denote equivalent models in terms of parameters to be estimated in the Discovery Phase.}
\label{fig:partial_order_mclust}
 \end{figure}
\item
  \emph{Step 3 - Constrained Maximization}: the parameter estimates are updated,
  based on the non-discarded observations and the current estimates for the unknown labels. Due to the constraint $\left( \sum_{g=1}^G\tau_g+\sum_{h=G+1}^E\tau_h\right)=1$, the mixing proportions are updated as follows using the Lagrange multipliers method (details can be found in the online supplementary material):
  \begin{align*}
\hat{\tau}_g^{(k+1)}=
\begin{cases} 
   \bar{\tau}_g \left( 1- \sum_{h=G+1}^E m^{*}_h \right) & g=1,\ldots, G \\
  \frac{\sum_{m=1}^{M} \varphi(\mathbf{y}_m^*) \hat{z}_{mg}^{*^{(k+1)}}}{\lceil M^*(1-\alpha_u)\rceil} & g=G+1,\ldots, E \\
  \end{cases}\\  
      \end{align*}
where 
\[m_h^{*}=\frac{\sum_{m=1}^{M} \varphi(\mathbf{y}_m^*) \hat{z}_{mh}^{*^{(k+1)}}}{\lceil M^*(1-\alpha_u)\rceil}.\]
In words, the proportions for the $G$ known classes computed in the learning phase are renormalized according to the proportions of the $H$ new groups. The estimate update for the mean vectors of the hidden classes reads:      
    \begin{equation*}
  \hat{\boldsymbol{\mu}}_h^{(k+1)}=\frac{\sum_{m=1}^{M^*}\varphi(\mathbf{y}^*_m)\hat{z}_{mh}^{*{(k+1)}}\mathbf{y}^*_m}{\sum_{m=1}^{M^*}\varphi(\mathbf{y}^*_m)\hat{z}_{mh}^{*{(k+1)}}}\:\:\:\:\: h=G+1,\ldots, E.
    \end{equation*}
    Estimation of $\boldsymbol{\Sigma}_h$, $h=G+1,\ldots,E$ depends on the selected patterned structure conditioning on the one estimated in the learning phase. More specifically, the parsimonious Gaussian models define a partial order in terms of model complexity. We allow for constraints relaxation when estimating the covariance matrices for the $H$ hidden classes, moving from left to right in the graph of Figure \ref{fig:partial_order_mclust}. While a full account on the inductive covariance matrices estimation is postponed to Appendix A, a simple example is reported here to clarify the procedure. Imagine to have selected a VEE model in the Learning
Phase: $\bar{\boldsymbol{\Sigma}}_g=\bar{\lambda}_g\bar{\mathbf{D}}\bar{\mathbf{A}}\bar{\mathbf{D}}^{'}$, $g=1,\ldots,G$. Due to the Inductive approach, the first $G$ covariance matrices need to be kept fixed with their volume already free to vary across components, so that only VEE, VVE, VEV and VVV models can be selected. Considering, for instance, a VEV  model (i.e., equal shape across components) in the discovery phase, the estimates for $\boldsymbol{\Sigma}_h$, $h=G+1, \ldots, E$ will be:
\begin{equation*}
\hat{\boldsymbol{\Sigma}}_h^{(k+1)} = \hat{\lambda}^{(k+1)}_h\hat{\mathbf{D}}^{(k+1)}_h\bar{\mathbf{A}}\hat{\mathbf{D}}^{(k+1)'}_h
\end{equation*}
where the estimate for the shape $\bar{\boldsymbol{A}}$ comes from the learning phase, while $\hat{\lambda}^{(k+1)}_h$ and $\hat{\mathbf{D}}^{(k+1)}_h$ respectively denote the inductive estimation for the volume and orientation of the $h$-th new class. Closed form solutions are obtained for all 14 models, depending on the parsimonious structure selected in the Learning Phase, under the eigenvalue restriction: details are reported in Appendix A. 
\item

  \emph{Step 4 - Convergence of the EM algorithm}: if convergence has not been reached (see Section \ref{sec:convergence}), set \(k=k+1\) and repeat steps 1-4.
 \end{itemize}  
\end{itemize}

Notice that the EM algorithm is solely based on the augmented test units for estimating parameters of the hidden classes. That is, if $E=G$ no extra parameters will be estimated in the discovery phase and the inductive approach will reduce to a fully-supervised classification method.

The final output from the learning phase is a set of parameters $\{\bar{\tau}_g,\bar{\boldsymbol{\mu}}_g,\bar{\boldsymbol{\Sigma}}_g\}$, $g=1,\ldots,G$ for the known classes and values for the indicator function \(\zeta(\cdot)\) where \(\zeta(\mathbf{x}_n)=0\) identify $\mathbf{x}_n$ as an outlying observation. The final output from the discovery phase is a set of parameters $\{\hat{\boldsymbol{\mu}}_h,\hat{\boldsymbol{\Sigma}}_h\}$, $h=G+1,\ldots,E$, for the hidden classes together with an update for the mixing proportion $\hat{\tau}_g$, $g=1,\ldots,E$ and values for the indicator functions \(\varphi(\cdot)\) where \(\varphi(\mathbf{y}^{*}_m)=0\) identify $\mathbf{y}^*_m$ as an outlying observations. Likewise for the transductive approach, the estimated values \(\hat{z}^*_{mg}\) provide a classification for the unlabelled observations \(\mathbf{y}^*_m\), assigning them to one of the known or hidden classes.

R \citep{RCoreTeam2018} source code implementing the EM algorithms under the transductive and inductive approaches is available at \\https://github.com/AndreaCappozzo/raedda. A dedicated R package is currently under development.
\subsection{Mathematical properties of robust estimation methods}
Robust inferential procedures via trimming and constraints are not mere heuristics that protect parameter estimates from the bias introduced by contaminated samples. They are based on theoretical results ensuring the existence of the solution in both the sample and the population problem. In addition, consistency of the sample solution to the population one has been proven under very mild conditions on the underlying distribution \citep{Garcia-Escudero2015}. We contribute to the theory of robust estimation by proving the monotonicity of the algorithms through the following proposition:\\ \\
\textbf{Proposition 1:} The EM algorithms described in Section \ref{sec:EM_transductive}  and \ref{sec:EM_inductive} imply
\[\ell_{trim}(\hat{\boldsymbol{\Theta}}^{(k+1)}| \mathbf{X}, \mathbf{Y},\mathbf{\text{l}}) \geq \ell_{trim}(\hat{\boldsymbol{\Theta}}^{(k)}| \mathbf{X}, \mathbf{Y},\mathbf{\text{l}})\]
for the objective function \eqref{trim_ll} in the transductive approach and \[\ell_{trim}(\hat{\boldsymbol{\Theta}}^{(k+1)}| \mathbf{Y}^*) \geq \ell_{trim}(\hat{\boldsymbol{\Theta}}^{(k)}| \mathbf{Y}^*)\]
for the objective function \eqref{trim_ll_discovery} in the discovery phase of the inductive approach, at any $k$, respectively.\\

The proof is reported in the online supplementary material, in which more details on the computing times of the proposed algorithms are also discussed.

\subsection{Convergence Criterion} \label{sec:convergence}
The convergence for both transductive and inductive approaches is assessed via the Aitken acceleration \citep{Aitken1926, McNicholas2010b}:
\begin{equation}
a^{(k)}=\frac{\ell_{trim}^{(k+1)}-\ell_{trim}^{(k)}}{\ell_{trim}^{(k)}-\ell_{trim}^{(k-1)}}
\end{equation}
where $\ell_{trim}^{(k)}$ is the trimmed observed data log-likelihood from iteration $k$: equation \eqref{trim_ll} and \eqref{trim_ll_discovery} for the transductive and the inductive approach, respectively.

 The asymptotic estimate of the trimmed log-likelihood at iteration $k$ is given by \citep{Bohning94}:
\begin{equation}
\ell_{\infty_{trim}}^{(k)} = \ell_{trim}^{(k)} + \frac{1}{1-a^{(k)}}\left(\ell_{trim}^{(k+1)}-\ell_{trim}^{(k)}\right).
\end{equation}
The EM algorithm is considered to have converged when $|\ell_{\infty_{trim}}^{(k)}-\ell_{trim}^{(k)}|<\varepsilon$; a value of $\varepsilon=10^{-5}$ has been chosen for the experiments reported in the next Sections.
\begin{table*}
\centering
\caption{Nomenclature and number of free parameters to be estimated for the variance covariance matrices, under the 14 patterned structures of \cite{Banfield1993} and \cite{Celeux1995}. $\gamma$ denotes the number of parameters related to the orthogonal rotation and $\delta$ the number of parameters related to the eigenvalues, for both transductive and inductive approach (discovery phase). The last column indicates whether the eigenvalue-ratio (ER) constraint is required. The learning phase of the inductive approach possesses the number of parameters indicated for the transductive approach, with $E$ replaced by $G$.}
\label{tab:TBIC_table}
\begin{tabular}{lccccc}
\hline\noalign{\smallskip}
 Model &$\gamma_{transductive}$& $\delta_{transductive}$&$\gamma_{inductive}$& $\delta_{inductive}$ & ER \\
 \noalign{\smallskip}\hline\noalign{\smallskip}
EII & $0$ & $1$ & $0$ & $0$ & Not Required \\ 
  VII & $0$ & $E$ & $0$ & $H$ & Required \\ 
  EEI & $0$ & $p$ & $0$ & $0$ & Not Required \\ 
  VEI & $0$ & $E + p - 1$ & $0$ & $H$ & Required \\ 
  EVI & $0$ & $Ep - (E - 1)$ & $0$ & $Hp - H$ & Required \\ 
  VVI & $0$ & $Ep$ & $0$ & $Hp$ & Required \\ 
  EEE & $p(p - 1)/2$ & $p$ & $0$ & $0$ & Not Required \\ 
  VEE & $p(p - 1)/2$ & $E + p - 1$ & $0$ & $H$ & Required \\ 
  EVE & $p(p - 1)/2$ & $Ep - (E - 1)$ & $0$ & $Hp - H$ & Required \\
  EEV & $Ep(p - 1)/2$ & $p$ & $Hp(p - 1)/2$ & $0$ & Not Required \\ 
  VVE & $p(p - 1)/2$ & $Ep$ & $0$ & $Hp$ & Required \\  
  VEV & $Ep(p - 1)/2$ & $E + p - 1$ & $Hp(p - 1)/2$ & $H$ & Required \\ 
  EVV & $Ep(p - 1)/2$ & $Ep - (E - 1)$ & $Hp(p - 1)/2$ & $Hp - H$ & Required \\ 
  VVV & $Ep(p - 1)/2$ & $Ep$ & $Hp(p - 1)/2$ & $Hp$ & Required \\
\noalign{\smallskip}\hline
\end{tabular}
\end{table*}
\subsection{Model Selection: determining the covariance structure and the number of components} \label{sec:TBIC}
A robust likelihood-based criterion is employed for choosing the number of hidden classes, the best model
among the 14 patterned covariance structures depicted in Figure \ref{fig:mclust_model} and a reasonable
value for the constraint $c$ in \eqref{eigen_contr}. Particularly, in our context, the problem of estimating the number of hidden classes corresponds to setting the number of components in a finite Gaussian mixture model (see for example \cite{Mclachlan2014} for a discussion on the topic). The general form of the robust information criterion is:
\begin{equation} \label{RBIC}
RBIC = 2 \ell_{trim}(\hat{\boldsymbol{\tau}}, \hat{\boldsymbol{\mu}}, \hat{\boldsymbol{\Sigma}}) - v_{XXX}^c \log\left(n^*\right)
\end{equation}
where $\ell_{trim}(\hat{\boldsymbol{\tau}}, \hat{\boldsymbol{\mu}}, \hat{\boldsymbol{\Sigma}})$ denotes the maximized trimmed observed data log-likelihood under either the transductive or inductive approach: equation \eqref{trim_ll} and \eqref{trim_ll_discovery}, respectively. The total number of observations $n^*$ employed in the estimation procedure is:
\[n^*= \begin{cases}
\lceil N(1-\alpha_{l})\rceil + \lceil M(1-\alpha_{u})\rceil & \text{Transductive EM}\\
\lceil M^*(1-\alpha_{u})\rceil & \text{Inductive EM} 
\end{cases}\]
In \eqref{RBIC}, the penalty term $v_{XXX}^c$ accounts for the number of parameters to be estimated. It depends on the estimation procedure (either transductive or inductive), the chosen patterned covariance structure (identified by the three letters subscript $XXX$, where $X$ can be either $E$, $V$ or $I$, like in Figure \ref{fig:mclust_model}) and the value for the constraint $c$: 
\begin{equation}
v_{XXX}^c = \kappa + \gamma + (\delta-1)\left(1-\frac{1}{c}\right)+1.
\end{equation}
$\kappa$ is the number of parameters related to the mixing proportions and the mean vectors: $\kappa=Ep+(E-1)$ in the transductive setting and $\kappa=Hp+(E-1)$ for the discovery phase in the inductive approach. $\gamma$ and $\delta$ denote, respectively, the number of free parameters related to the orthogonal rotation and to the eigenvalues for the estimated covariance matrices. Their values, for the two approaches and the different patterned structure, are reported in Table \ref{tab:TBIC_table}.

The robust information criterion in \eqref{RBIC} is an adaptation of the complexity-penalized likelihood approach introduced in \cite{Cerioli2018} that here also accounts for the trimming levels and patterned structures. Note that, when $c\rightarrow +\infty$ and $\alpha_l=\alpha_u=0$, \eqref{RBIC} reduces to the well-known Bayesian Information Criterion \citep{Schwarz1978}. 

Even though the RBIC in \eqref{RBIC} is shown to work well in all the simulated experiments of Section \ref{sec:sim_study} and in the microbiome analyses of Section \ref{sec:application}, 
a more general consideration on the usage of trimming criteria to perform model selection in robust mixture learning is in order. Firstly proposed by \cite{Neykov2007}, the authors asserted that the trimmed BIC (TBIC) could be employed for robustly assessing the number of mixture components and the percentage of contamination in the data. Since its first introduction, trimming criteria have been extensively employed in the literature for providing/suggesting a general way to perform robust model selection \citep{Garcia-Escudero2010, Gallegos2010, garcia2016joint, Garcia-Escudero2017, Li2016, Cerioli2018b}.
Quite naturally, the rationale behind such criteria
stems from the need of defining a model selection procedure whose output should result close to that obtained by standard methods on the genuine part of the data only. Indeed, robustly estimated parameters are not sufficient to provide reliable model selection if the maximized likelihood is evaluated on the entire dataset: noisy units contribute to the value of standard criteria and their effect, albeit small, could affect the overall behavior. An example of such undesirable outcome, in a weighted likelihood framework, is reported in Section 5 of \cite{Greco2019}. Therefore, it is reasonable to perform model comparison within the subset of genuine units only, where the subset size is determined by the trimming levels and the definition of ``genuine'' is in accordance with the estimated model. That is, as correctly emphasized by an anonymous reviewer, RBIC indexes could be based on different subsets of observations when considering different parameterizations. This shall not be viewed as a criterion drawback, since the RBIC precisely aims at identifying, in a data-driven fashion, the model that better fits the uncontaminated subgroup.

All in all, even though a formal theory corroborating trimming criteria is still missing in the literature and model-selection consistency guarantees is yet to be derived, RBIC provides a well-established and powerful technique for comparing models conditioning on the same trimming fractions, as performed in the paper.

\subsection{On the role of the eigenvalue restrictions} \label{sec:eigen_ratio}

Extensive literature has been devoted to studying the appearance of the so-called \textit{degenerate solutions} that may be provided by the EM algorithm when fitting a finite mixture to a set of data \citep{Peel2000,BIERNACKI2007, Ingrassia2011}. This is due to the likelihood function itself, rather than being a shortcoming of the EM procedure: it is easy to show that for elliptical mixture models with unrestricted covariance matrices the associated likelihood is unbounded \citep{day1969estimating}. An even more subtle problem, at least from a practitioner perspective, is the appearance of solutions that are not exactly degenerate, but they can be regarded as spurious since they lie very close to the boundary of the parameter space, namely when a fitted component has a very small generalized variance \citep{Peel2000}. Such solutions correspond to situations in which a mixture component fits few data points almost lying in a lower-dimensional subspace. They often display a high likelihood value, whilst providing little insight in real-world applications. They mostly arise as a result of modeling a localized random pattern rather than a true underlying group. Many possible solutions have been proposed in the literature to tackle the problem, a comprehensive list of such references can be found in \cite{Garcia-Escudero}.

When employing mixture models for supervised learning and discriminant analysis there is actually no need in worrying about the appearance of spurious solutions, since the joint distribution of both observations and associated labels is directly available. The parameters estimation therefore reduces to estimate the within class mean vector and covariance matrix, without the need of any EM algorithm \citep{Fraley2002}. Nonetheless, adaptive learning is based on a partially unsupervised estimation, since hidden classes are sought in the test set without previous knowledge of their group structure extracted from the labelled set. Therefore, efficiently dealing with the possible appearance of spurious solutions becomes fundamental in our context, where the identification of a hidden class might just be the consequence of a spurious solution. 
For an extensive review of eigenvalues and constraints in mixture modeling, the interested reader is referred to \cite{Garcia-Escudero} and references therein.
\begin{figure}
\centering
\includegraphics[scale=0.4]{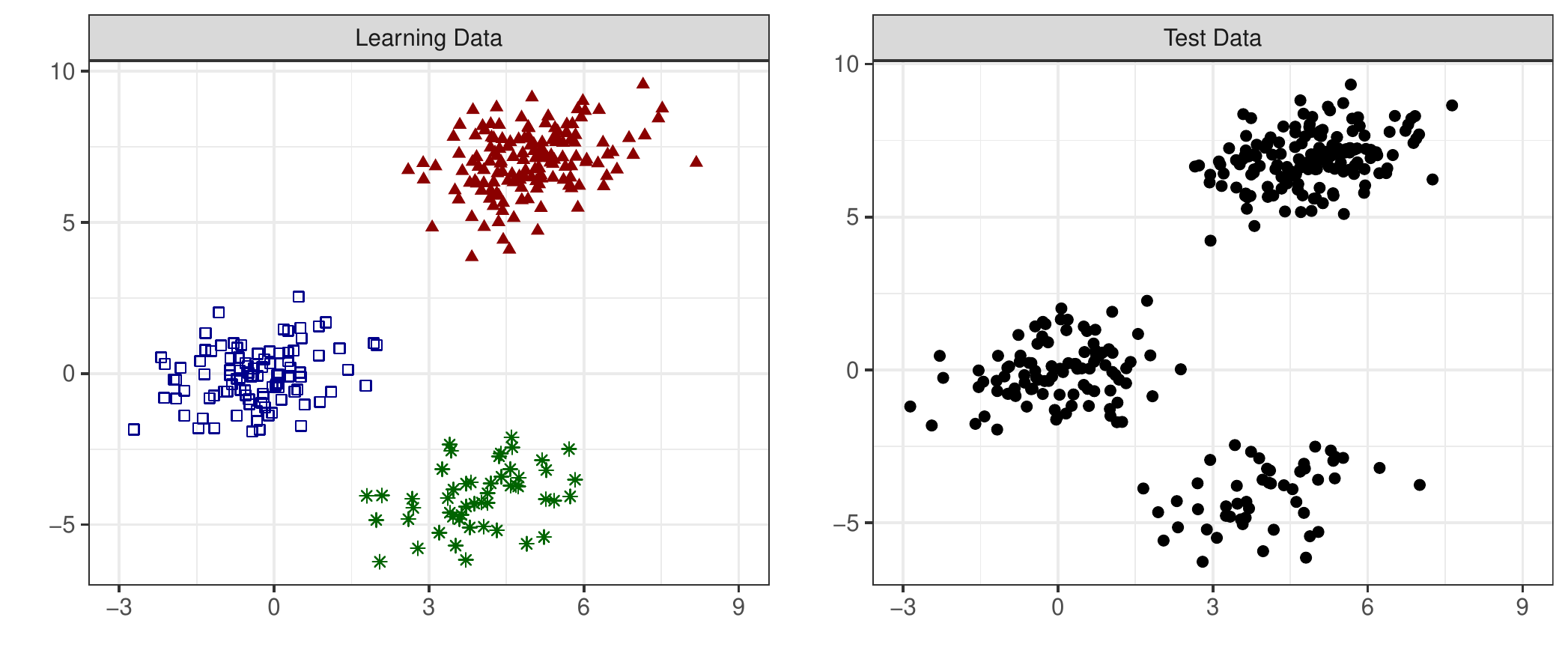}
\caption{Original learning problem, with a set of $N=300$ labelled observations and $M=300$ unlabelled observations generated from the same mixture of bivariate normal distributions with three components.}
\label{fig:sp_example}
\end{figure}

\begin{figure}
\centering
\includegraphics[scale=0.4]{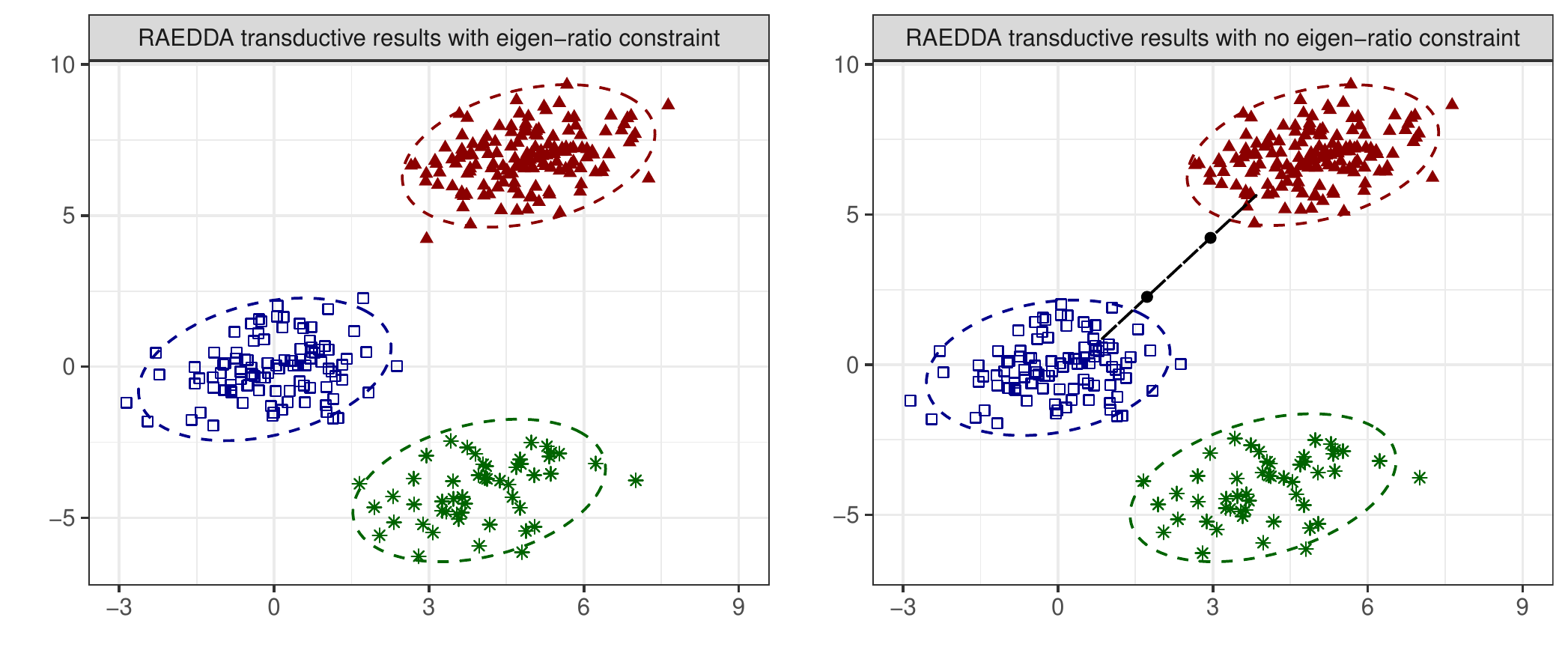}
\caption{The classification obtained for the best model in the test set, with two different values for the eigen-ratio constraint. In the unconstrained case the classification is based on a spurious solution, with a localized random pattern wrongly identified as a hidden class.}
\label{fig:sp_results}
\end{figure}

We now provide an illustrative example for underlying the importance of protecting the adaptive learner from spurious solutions, that may arise also in the simplest scenarios. Consider a data generating process given by a three components mixture of bivariate normal distributions ($E=G=3$ and $p=2$) with the following parameters:
\[ \boldsymbol{\tau}=(0.35, 0.15, 0.5)',\] \[\boldsymbol{\mu}_1=(0, 0)', \quad \boldsymbol{\mu}_2=(4, -4)', \quad \boldsymbol{\mu}_3=(5, 7)'\]
 \[ \boldsymbol{\Sigma}_1 =\boldsymbol{\Sigma}_2=\boldsymbol{\Sigma}_3= \begin{bmatrix}
    1       & 0.3\\
    0.3       & 1
    \end{bmatrix}\]
Figure \ref{fig:sp_example} graphically presents the learning problem, in which both the training and test sets contain $300$ data points. Clearly, even from a visual exploration, the test set does not contain any hidden group and we therefore expect that the model selection criterion defined in Section \ref{sec:TBIC} will choose a mixture of $E=3$ components as the best model for the problem at hand. Employing transductive estimation, the RAEDDA model is fitted to the data, with trimming levels set to $0$ for both labelled and unlabelled sets ($\alpha_l=\alpha_u=0$) and considering two different values for the eigen-ratio constraint: $c=10$ in the first case and $c=10^{10}$ in the second. That is, we set a not too restrictive constraint in the former model (notice that the true ratio between the biggest and smallest eigenvalues of $\boldsymbol{\Sigma}_g$, $g=1,\ldots,3$ is equal to 1.86) and we consider a virtually unconstrained estimation for the latter. The classification obtained for the best model in the test set, selected via the robust information criterion in \eqref{RBIC}, under the two different scenarios is reported in Figure \ref{fig:sp_results}.
The value for the maximized log-likelihood in the first scenario is equal to $-2257.279$, and it is equal to $-2186.615$ in the unconstrained case. With only 2 data points in the hidden group and $|\hat{\boldsymbol{\Sigma}}_4| < 10^{-10}$ we are clearly dealing with a spurious solution and not with a hidden class. Nonetheless, the appearance of spurious maxima even in this simple toy experiment casts light on how paramount it is to protect the estimates against this harmful possibility.  
\subsection{Further aspects}

Notice that the RAEDDA methodology is a generalization of several model-based classification methods, in particular:
\begin{itemize}
\item EDDA \citep{Bensmail1996} when only fitting the robust learning phase with $\alpha_l=0$.
\item REDDA \citep{Cappozzo2019b} when only fitting the robust learning phase with $\alpha_l>0$.
\item UPCLASS \citep{Dean2006} when fitting the transductive approach with $E=G$ and $\alpha_l=0$, $\alpha_u =0$
\item RUPCLASS \citep{Cappozzo2019b} when fitting the transductive approach with $E=G$ and $\alpha_l > 0$, $\alpha_u >0$.
\item AMDA transductive \citep{Bouveyron} when fitting the transductive approach with $E \geq G$ and $\alpha_l=0$, $\alpha_u =0$. Notice in addition that RAEDDA considers a broader class of learners employing patterned covariance structures.
\item AMDA inductive \citep{Bouveyron} when fitting the inductive approach with $\alpha_l=0$, $\alpha_u =0$. Also here the class of considered models is larger, thanks to the partial-order structure in the eigen-decomposition of the covariance matrices (see Figure \ref{fig:partial_order_mclust}).
\end{itemize}

\section{Simulation study} \label{sec:sim_study}
In this Section, we present a simulation study in which the performance of novelty detection methods is assessed when dealing with different combinations of data generating processes and contamination rates. For each scenario, an entire class is not present in the labelled units, and it thus needs to be discovered by the adaptive classifiers in the test set. The problem definition is therefore as follows: we aim at judging the performance of various methods in recovering the true partition under a semi-supervised framework, where the groups distribution is (approximately) Gaussian, allowing for a distribution-free noise structure, both in terms of label noise and outliers.

\subsection{Experimental Setup} \label{sim_setup}
The $E=3$ classes are generated via multivariate normal distributions of dimension $p=6$ with the following parameters:
 \[ \quad \boldsymbol{\mu}_1=(0, 8,0,0,0,0)', \quad \boldsymbol{\mu}_2=(8, 0,0,0,0,0)',\]
 \[\boldsymbol{\mu}_3=(-8, -8,0,0,0,0)', \quad \boldsymbol{\Sigma}_1 = \hbox{diag}(1,a,1,1,1,1),\] \[\boldsymbol{\Sigma}_2 = \hbox{diag}(b,c,1,1,1,1), \quad
    \boldsymbol{\Sigma}_3 = 
\left(\begin{array}{@{}c|c@{}}
  \begin{matrix}
  d & e \\
  e & f
  \end{matrix}
  & \boldsymbol{0} \\
\hline
  \boldsymbol{0} &
  \boldsymbol{I}
\end{array}\right)
    \]
We consider 5 different combinations of $(a,b,c,d,e, f)$:

\begin{itemize}
\item $(a,b,c,d,e, f)=(1,1,1,1,0,1)$, spherical groups with equal volumes (EII)
\item $(a,b,c,d,e, f)=(5,1,5,1,0,5)$, diagonal groups with equal covariance matrices (EEI)
\item $(a,b,c,d,e, f)=(5,5,1,3,-2,3)$, groups with equal volume, but varying shapes and orientations (EVV)
\item $(a,b,c,d,e, f)=(1,20,5,15,-10,15)$, groups with different volumes, shapes and orientations (VVV)
\item $(a,b,c,d,e, f)=(1,45,30,15,-10,15)$, groups with different volumes, shapes and orientations (VVV) but with two severe overlap
\end{itemize}
The afore-described data generating process has been introduced in \cite{Garcia-Escudero2008}: we adopt it here as it elegantly provides a well-defined set of resulting parsimonious covariance structures. In addition, two different group proportions are included:
 \begin{itemize}
 \item \texttt{equal}: $N_1=N_2=285$ and $M_1=M_2=M_3=360$
 \item \texttt{unequal}: $N_1=190, \: N_2=380$ and $M_1=210$, $M_2=430$, $M_3=60$
 \end{itemize}
where $N_g,\: g=1,2$ and $M_h, \: h=1,2,3$ denote the sample sizes for each group in the training and test sets, respectively. According to the notation introduced in Section \ref{sec:AMDA}, we observe $G=2$ classes in the training and $H=1$ extra class in the test set. Furthermore, we apply contamination adding both attribute and class noise as follows. A fixed number $Q_l$ and $Q_u$ of uniformly distributed outliers, having squared Mahalanobis distances from $\boldsymbol{\mu}_1, \boldsymbol{\mu}_2$ and $\boldsymbol{\mu}_3$ greater than $\chi^2_{6, 0.975}$, are respectively added to the labelled and unlabelled sets. Additionally, we assign a wrong label to $Q_l$ genuine units, randomly chosen in the training set. Four different contamination levels are considered, varying $Q_l$ and $Q_u$:
\begin{itemize}
\item No contamination: $Q_l=Q_u=0$,
\item Low contamination: $Q_l=10$ and $Q_u=40$,
\item Medium contamination: $Q_l=20$ and $Q_u=80$,
\item Strong contamination: $Q_l=30$ and $Q_u=120$.
\end{itemize} 
Notice that, for the \texttt{unequal} group proportion, the hidden class sample size is smaller than the total number of outlying units when medium and strong contamination is considered. A total of $B=1000$ Monte Carlo replications are generated for each combination of covariance structure, groups proportion and contamination rate. Results for the considered scenarios are reported in the next Section.      
\subsection{Simulation results}
\begin{figure*}
\begin{subfigure}{.5\textwidth}
  \centering
  \includegraphics[width=1\linewidth]{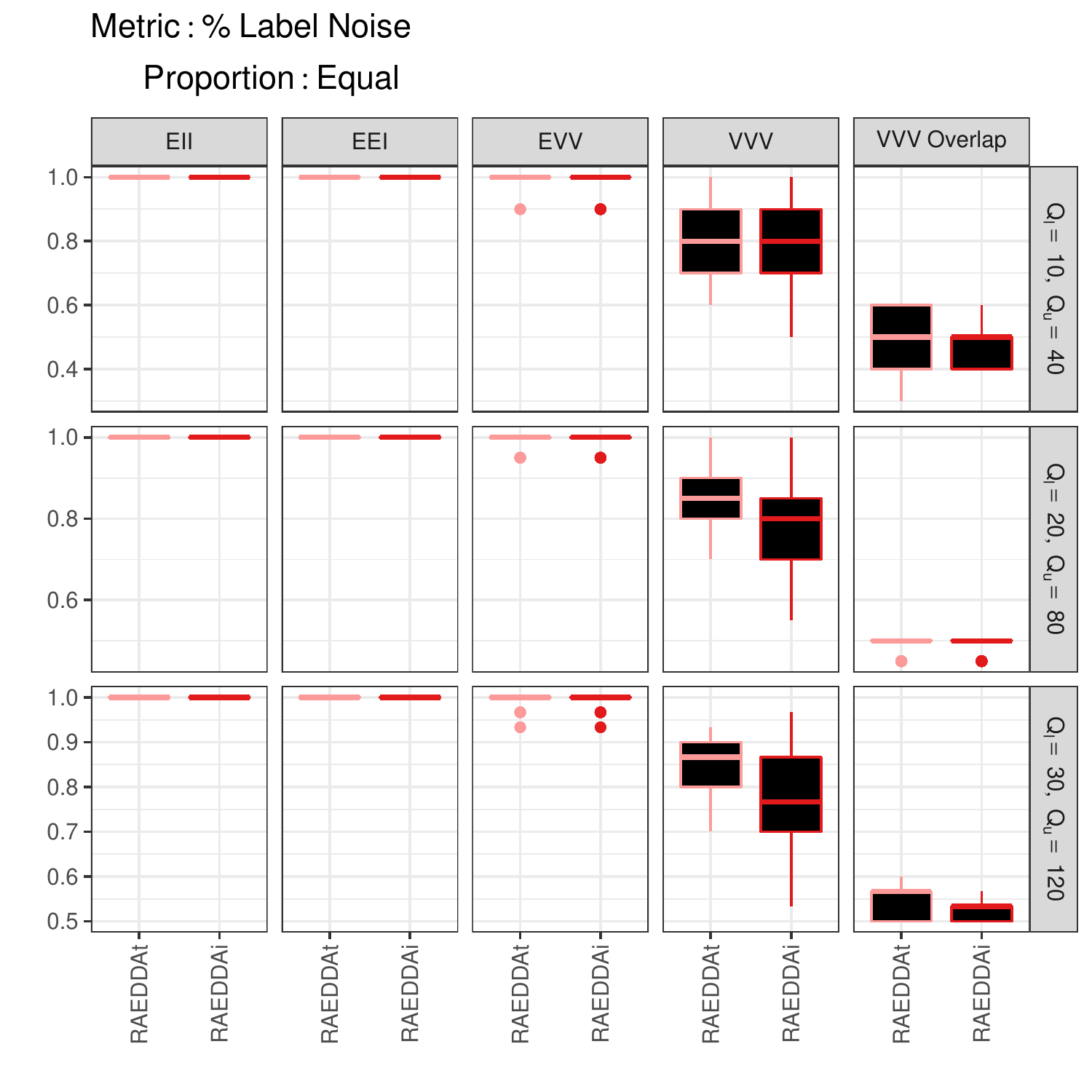}
\end{subfigure}%
\begin{subfigure}{.5\textwidth}
  \centering
  \includegraphics[width=1\linewidth]{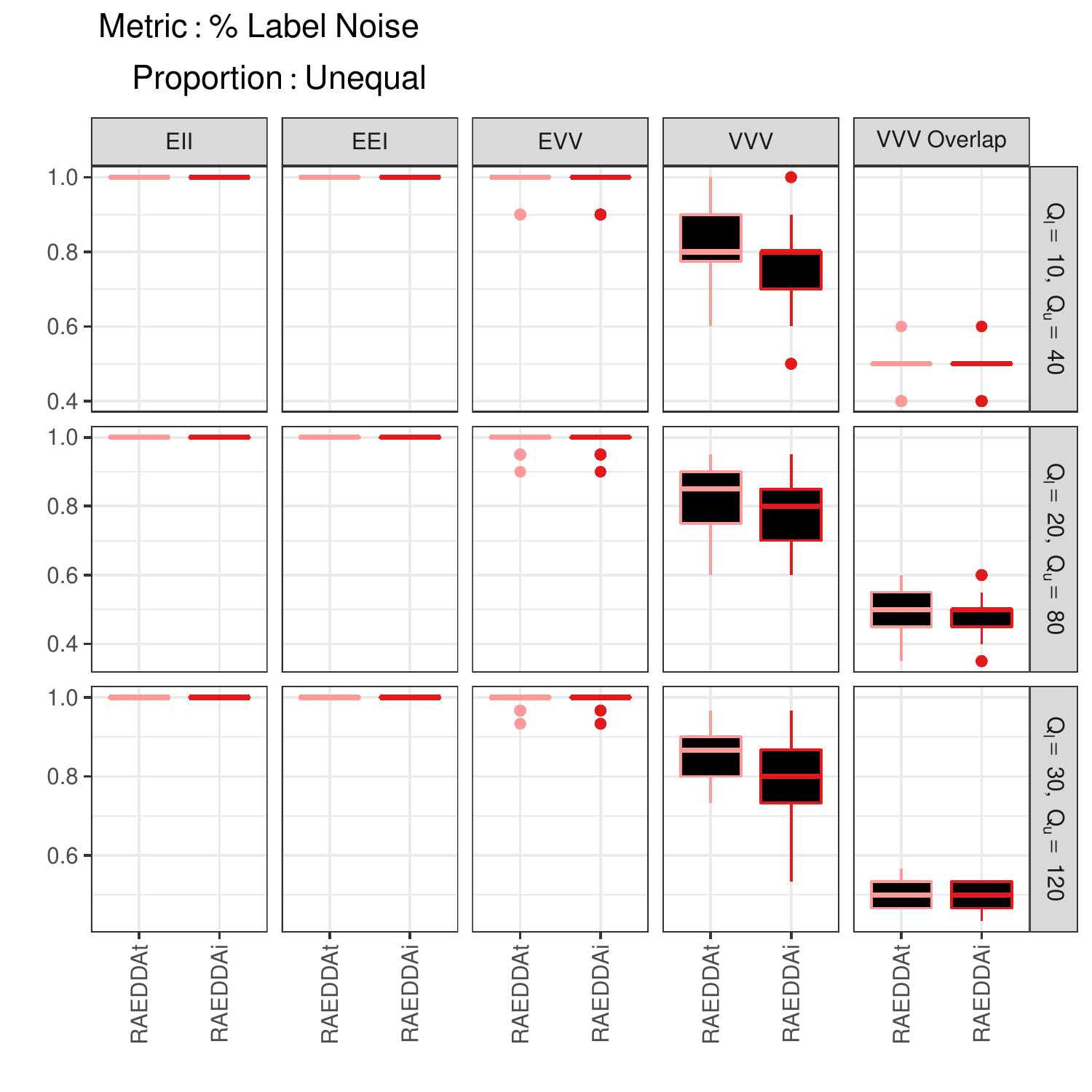}
\end{subfigure}
\begin{subfigure}{.5\textwidth}
  \centering
  \includegraphics[width=1\linewidth]{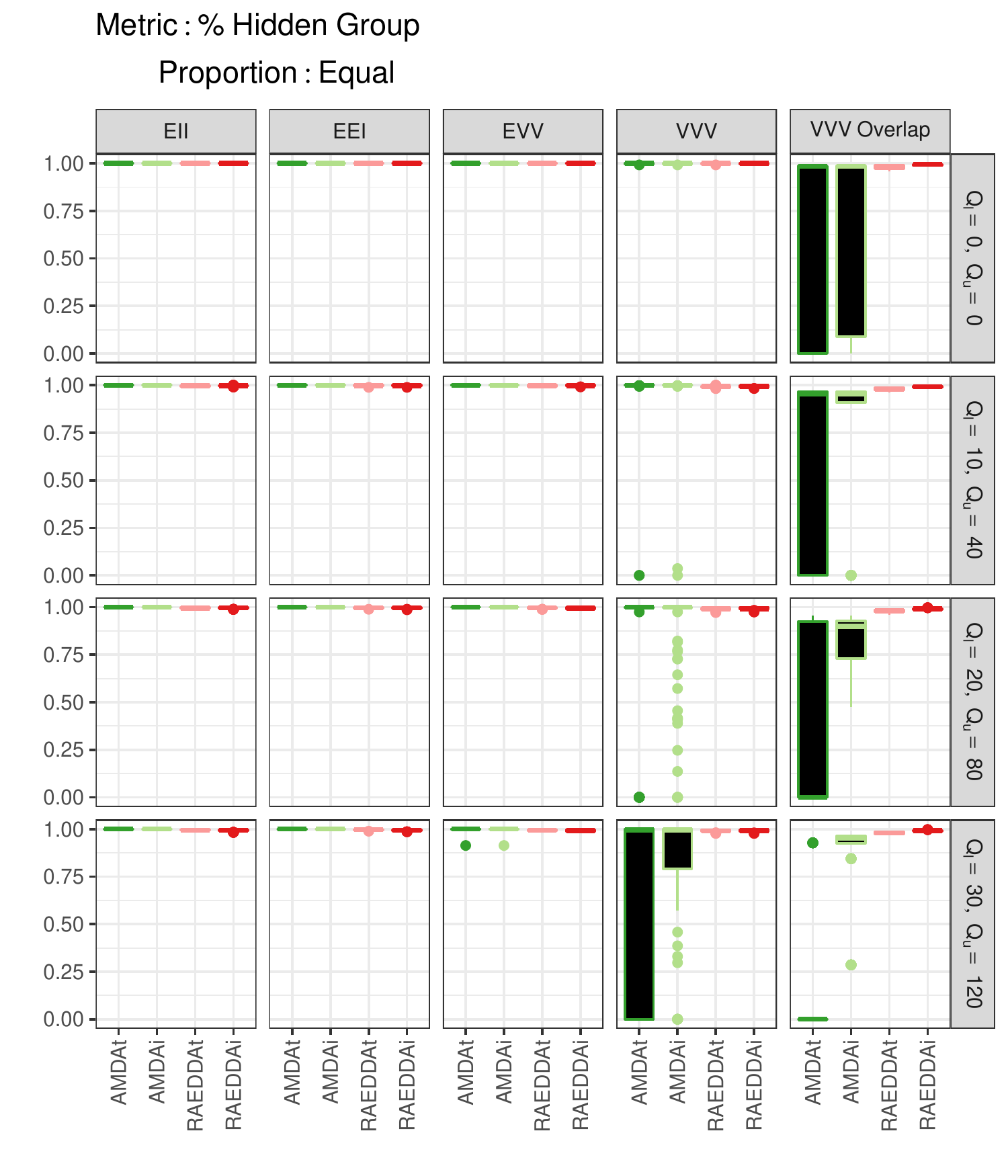}
\end{subfigure}%
\begin{subfigure}{.5\textwidth}
  \centering
  \includegraphics[width=1\linewidth]{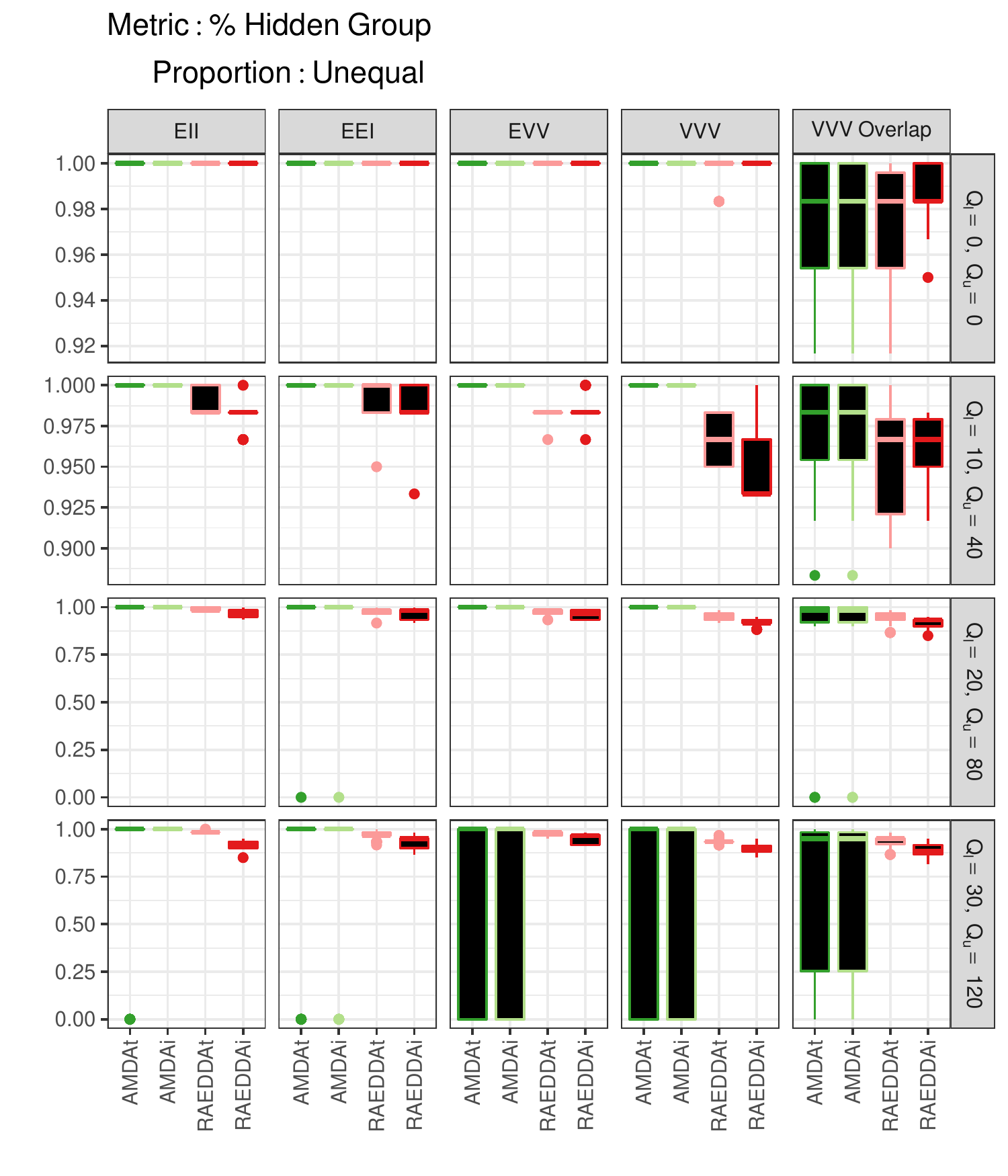}
\end{subfigure}
\caption{Box plots for \% Label Noise and \% Hidden Group metrics for $B = 1000$ Monte Carlo repetitions under different covariance structure, groups proportion and contamination rate.}
\label{fig:boxplot_sim_study_RAEDDA_1}
\end{figure*}

\begin{figure*}
\begin{subfigure}{.5\textwidth}
  \centering
  \includegraphics[width=1\linewidth]{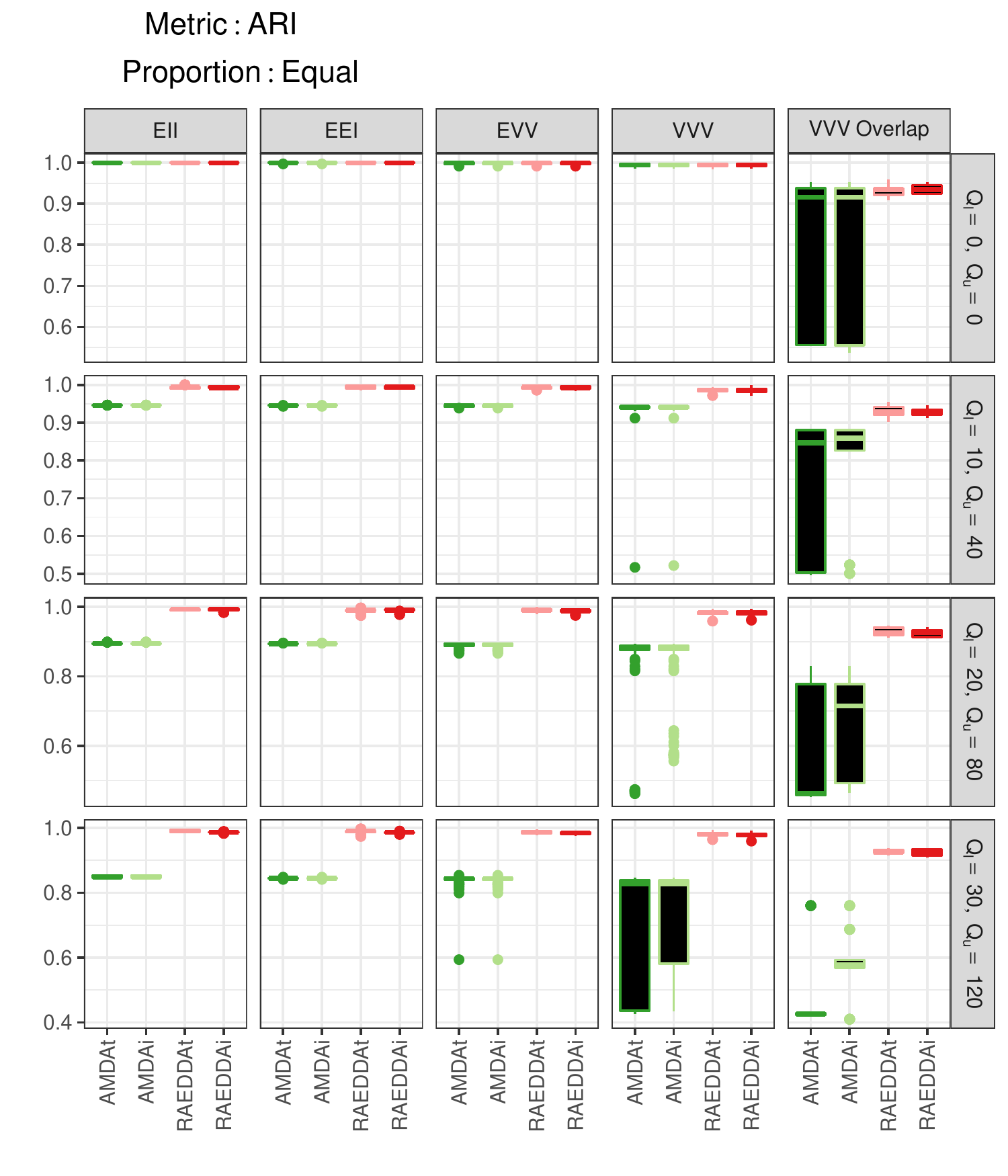}
\end{subfigure}%
\begin{subfigure}{.5\textwidth}
  \centering
  \includegraphics[width=1\linewidth]{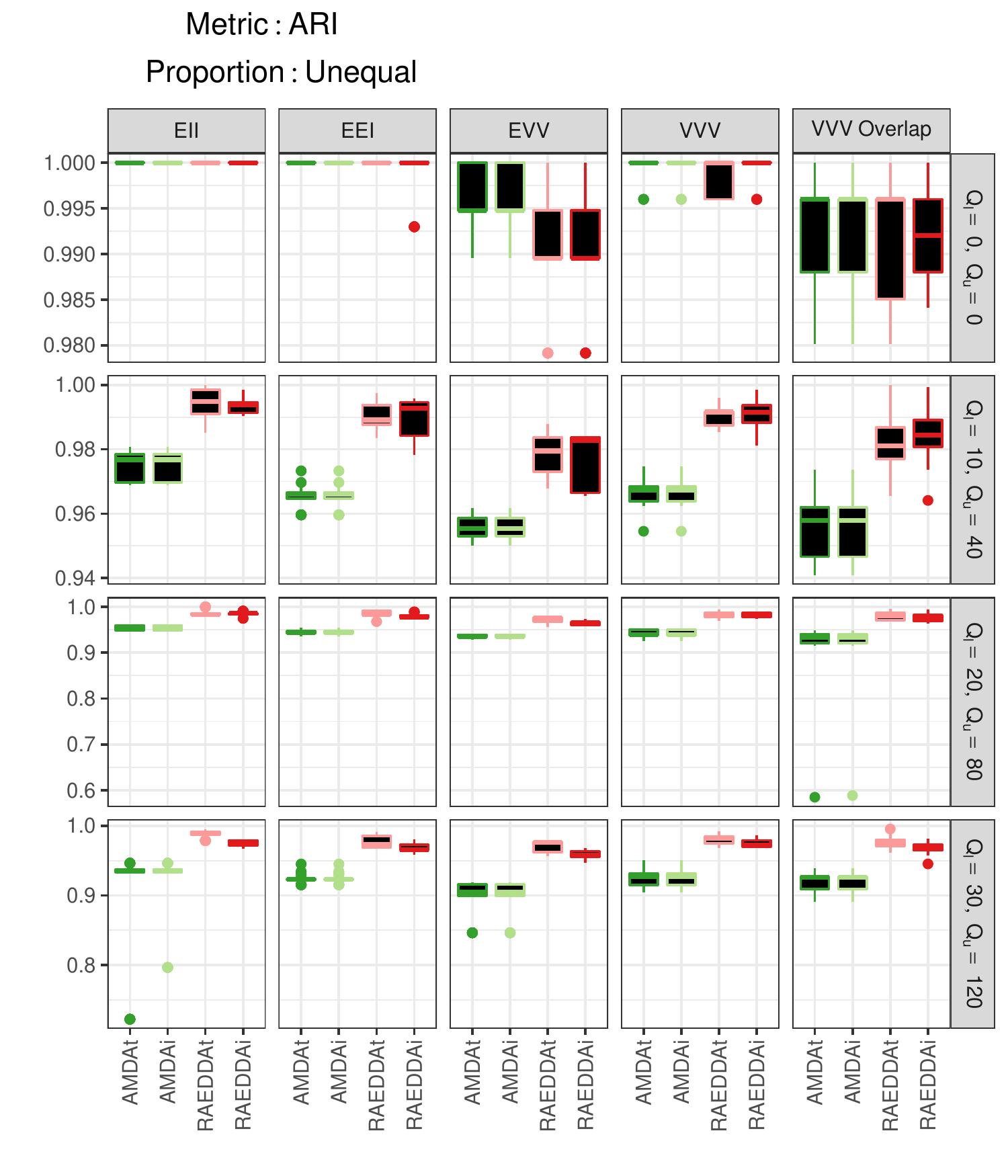}
\end{subfigure}
\begin{subfigure}{.5\textwidth}
  \centering
  \includegraphics[width=1\linewidth]{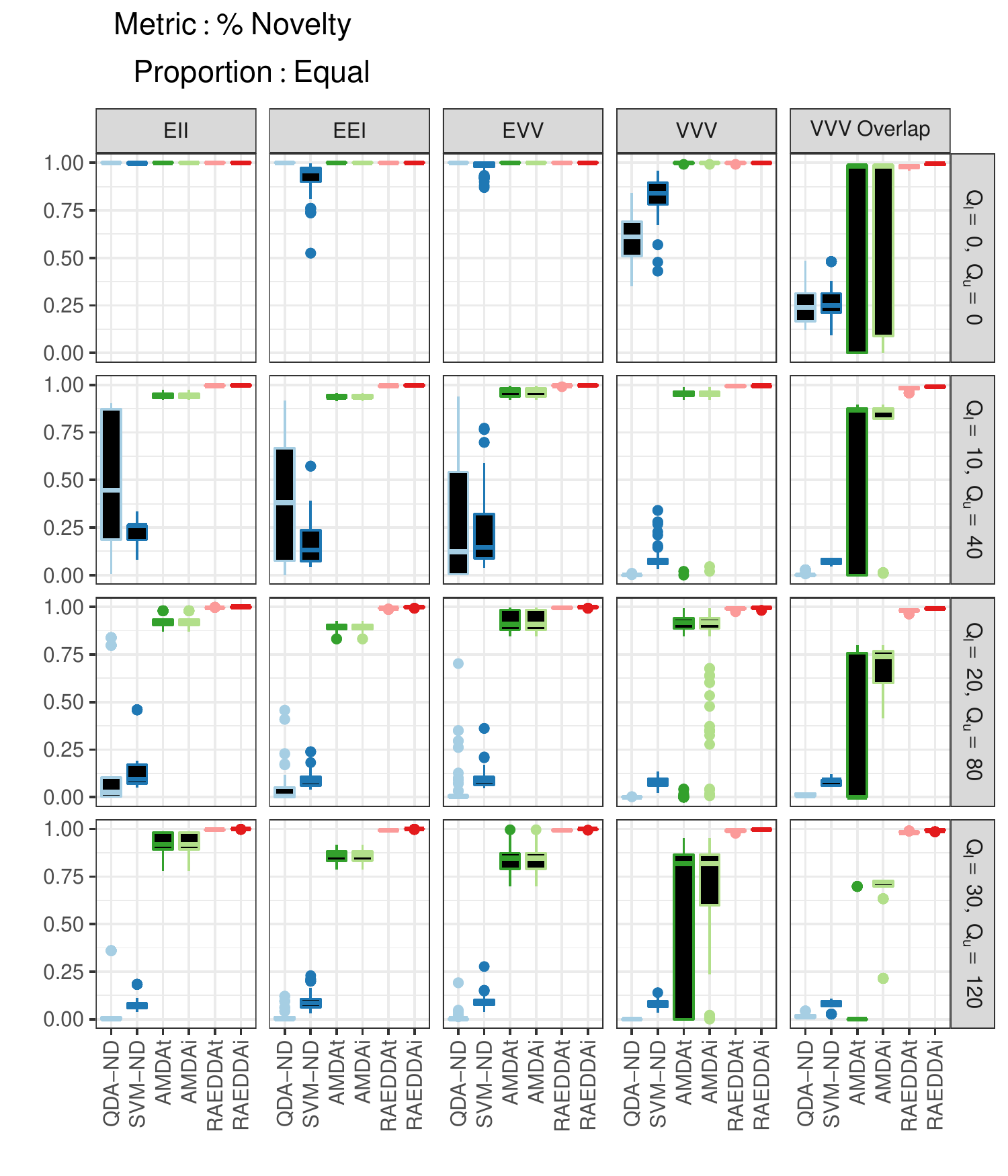}
\end{subfigure}%
\begin{subfigure}{.5\textwidth}
  \centering
  \includegraphics[width=1\linewidth]{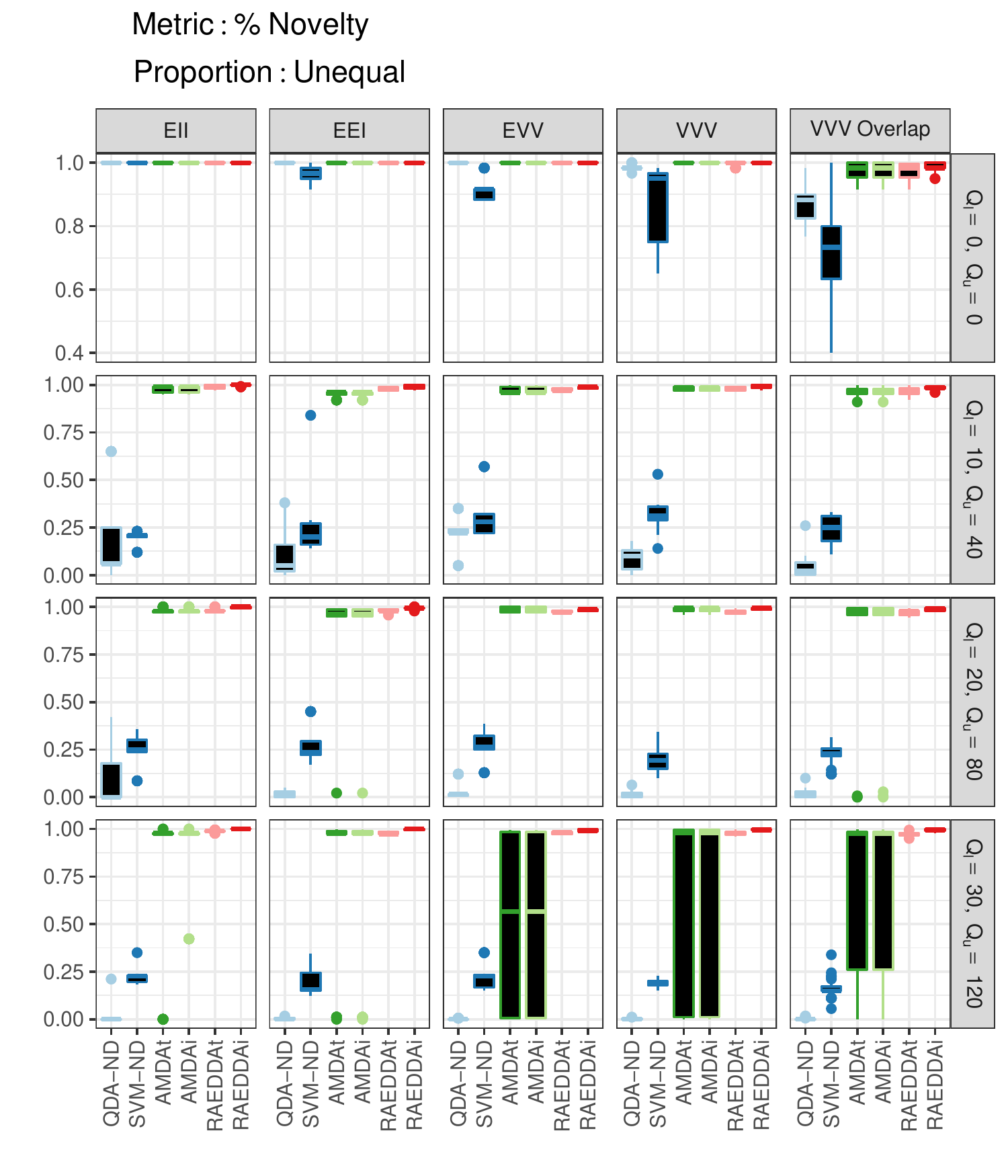}
\end{subfigure}
\caption{Box plots for ARI and \% Novelty metrics for $B = 1000$ Monte Carlo repetitions under different covariance structure, groups proportion and contamination rate.}
\label{fig:boxplot_sim_study_RAEDDA_2}
\end{figure*}

Given the simulation framework presented in the previous Section, we compare the performance of RAEDDA against the original AMDA model (denoting by RAEDDAt, AMDAt and RAEDDAi, AMDAi their transductive and inductive versions) and two popular novelty detection methods, namely Classifier Instability \citep{Tax1998} and Support Vector Method for novelty detection \citep{Scholkopf2000}, respectively denoted as QDA-ND and SVM-ND hereafter. 
For assessing the performance in terms of classification accuracy, outliers detection and hidden groups discovery for the competing methods, a set of 4 metrics is recorded at each replication of the simulation study:
\begin{itemize}
\item \% Label Noise: the proportion of $Q_l$ mislabelled units in the training set correctly identified as such by the RAEDDA model (for which the final value of the trimming function $\zeta(\cdot)$ is equal to $0$);
\item \% Hidden Group: the proportion of units in the test set belonging to the third group correctly assigned to a previously unseen class by AMDA and RAEDDA methods;
\item ARI: Adjusted Rand Index \citep{Rand1971}, measuring the similarity between the partition returned by a given method and the underlying true structure;
\item \% Novelty: the proportion of units in the test set belonging either to the third group or to the set of $Q_u$ outliers correctly identified by the novelty detection methods.
\end{itemize}
Box plots for the four metrics, resulting from the $B$ Monte Carlo repetitions under different covariance structure, groups proportion and contamination rate are reported in Figure \ref{fig:boxplot_sim_study_RAEDDA_1} and \ref{fig:boxplot_sim_study_RAEDDA_2}. 
\begin{figure*}
\begin{subfigure}{.195\textwidth}
  \centering
  \includegraphics[width=1\linewidth]{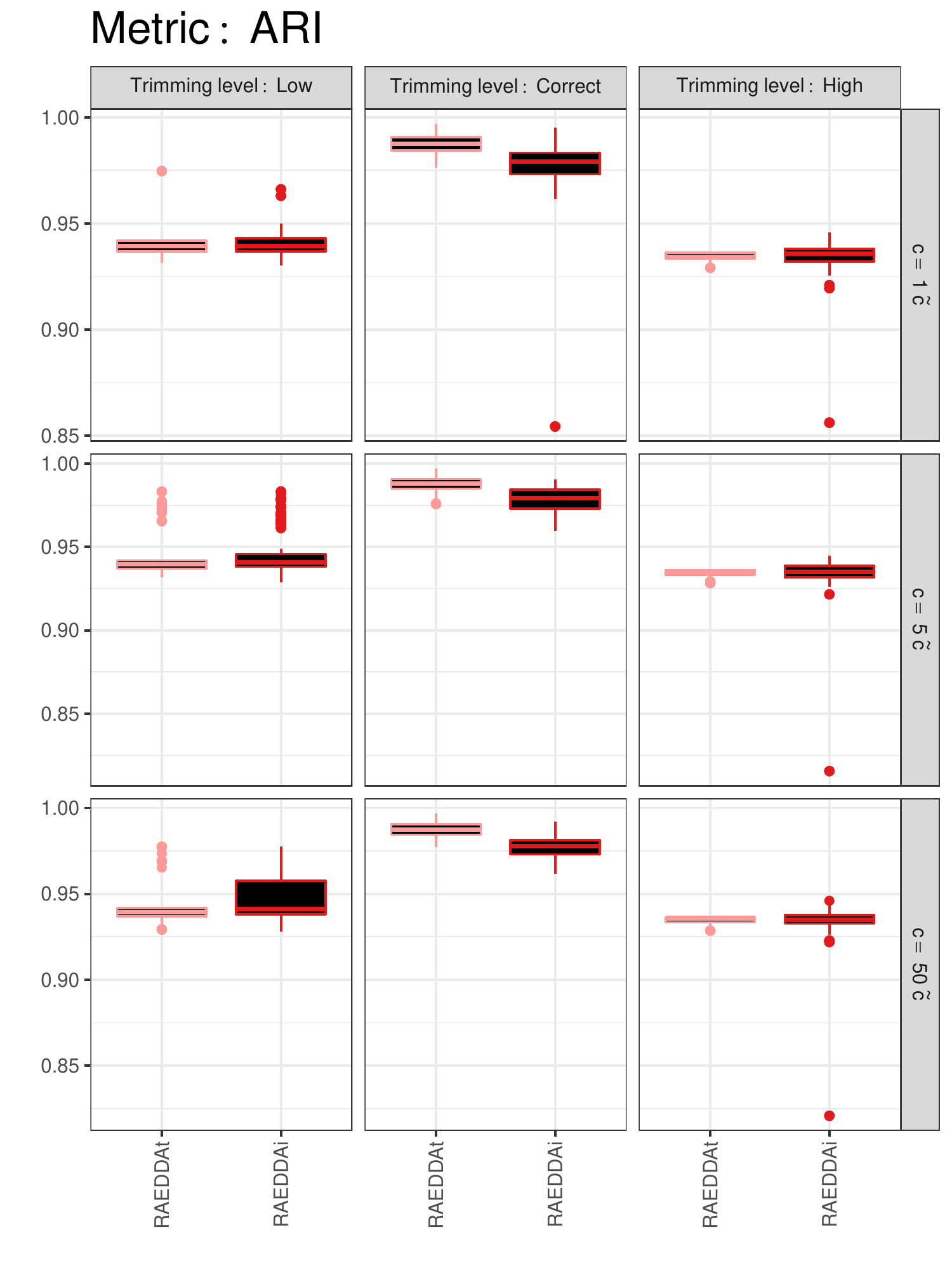}
\end{subfigure}%
\begin{subfigure}{.195\textwidth}
  \centering
  \includegraphics[width=1\linewidth]{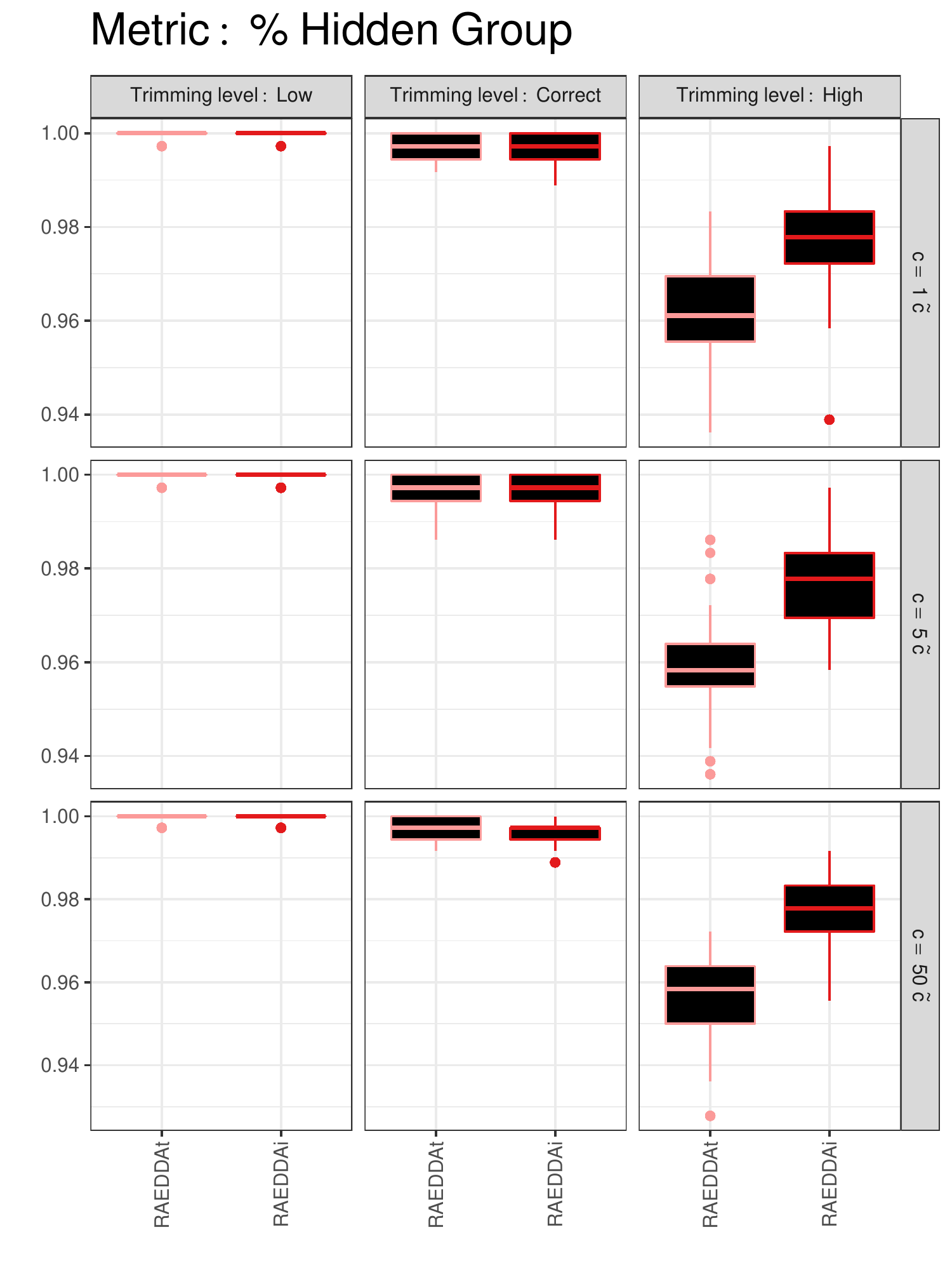}
\end{subfigure}
\begin{subfigure}{.195\textwidth}
  \centering
  \includegraphics[width=1\linewidth]{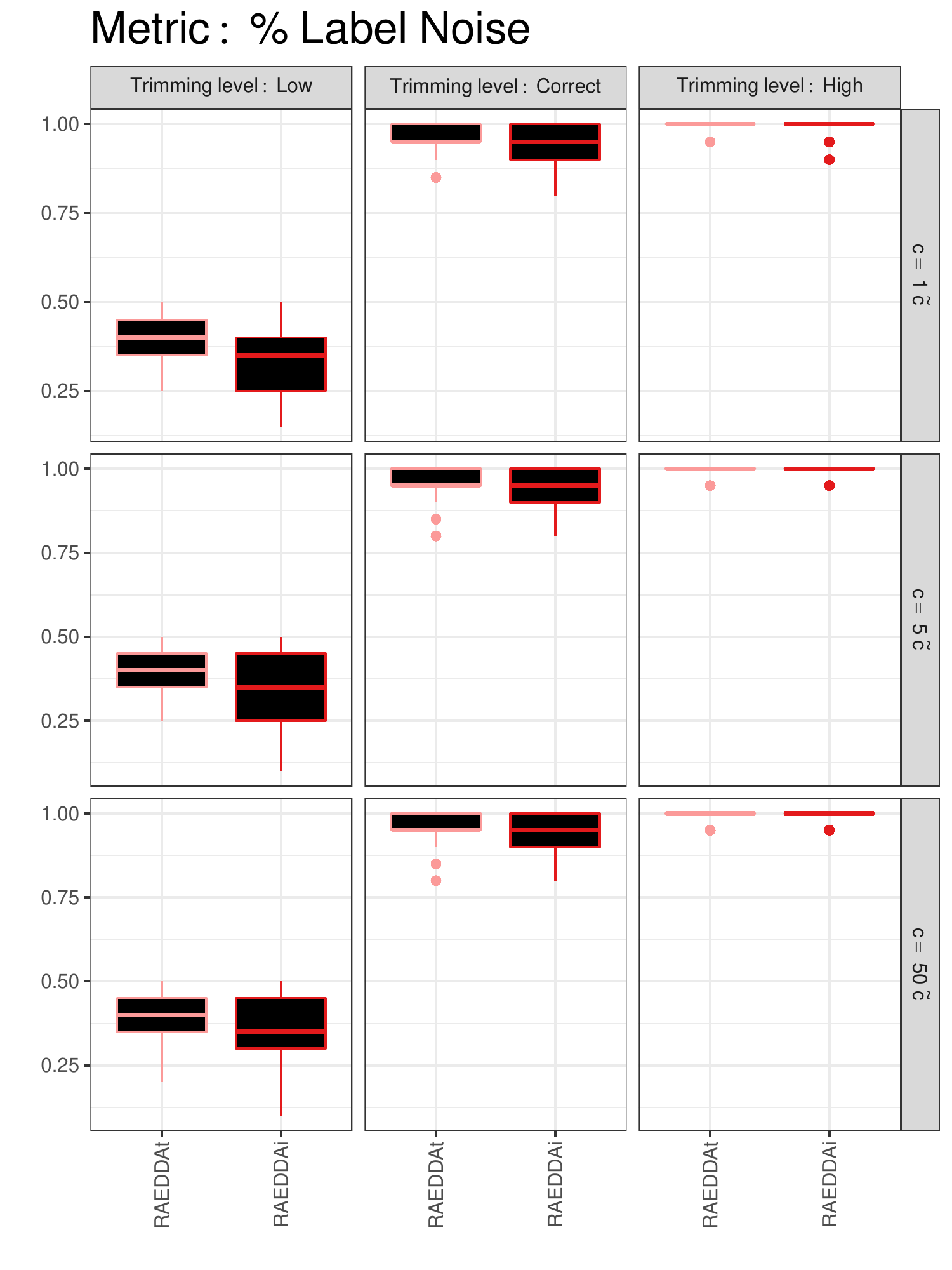}
\end{subfigure}%
\begin{subfigure}{.195\textwidth}
  \centering
  \includegraphics[width=1\linewidth]{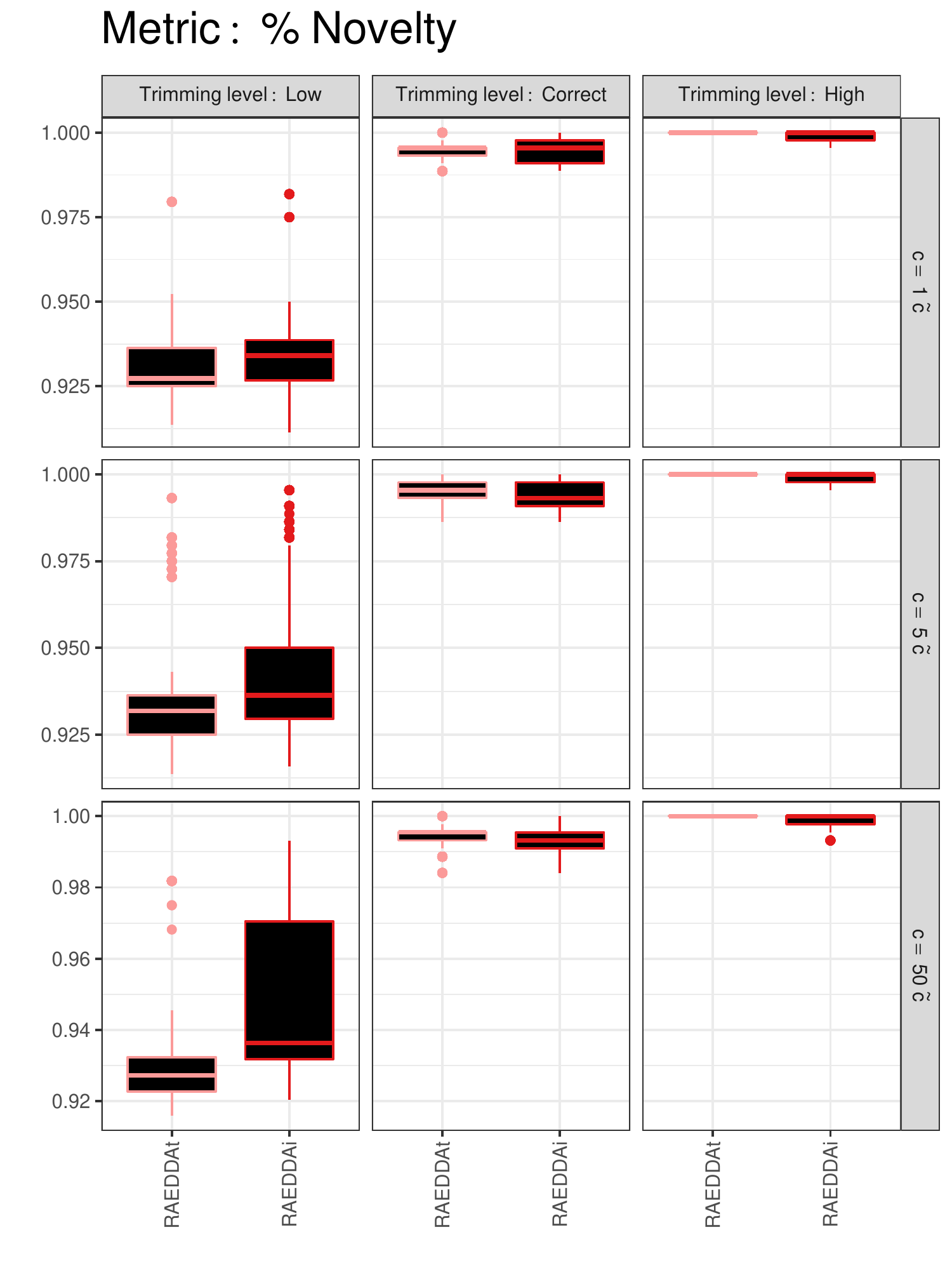}
\end{subfigure}
\begin{subfigure}{.195\textwidth}
  \centering
  \includegraphics[width=1\linewidth]{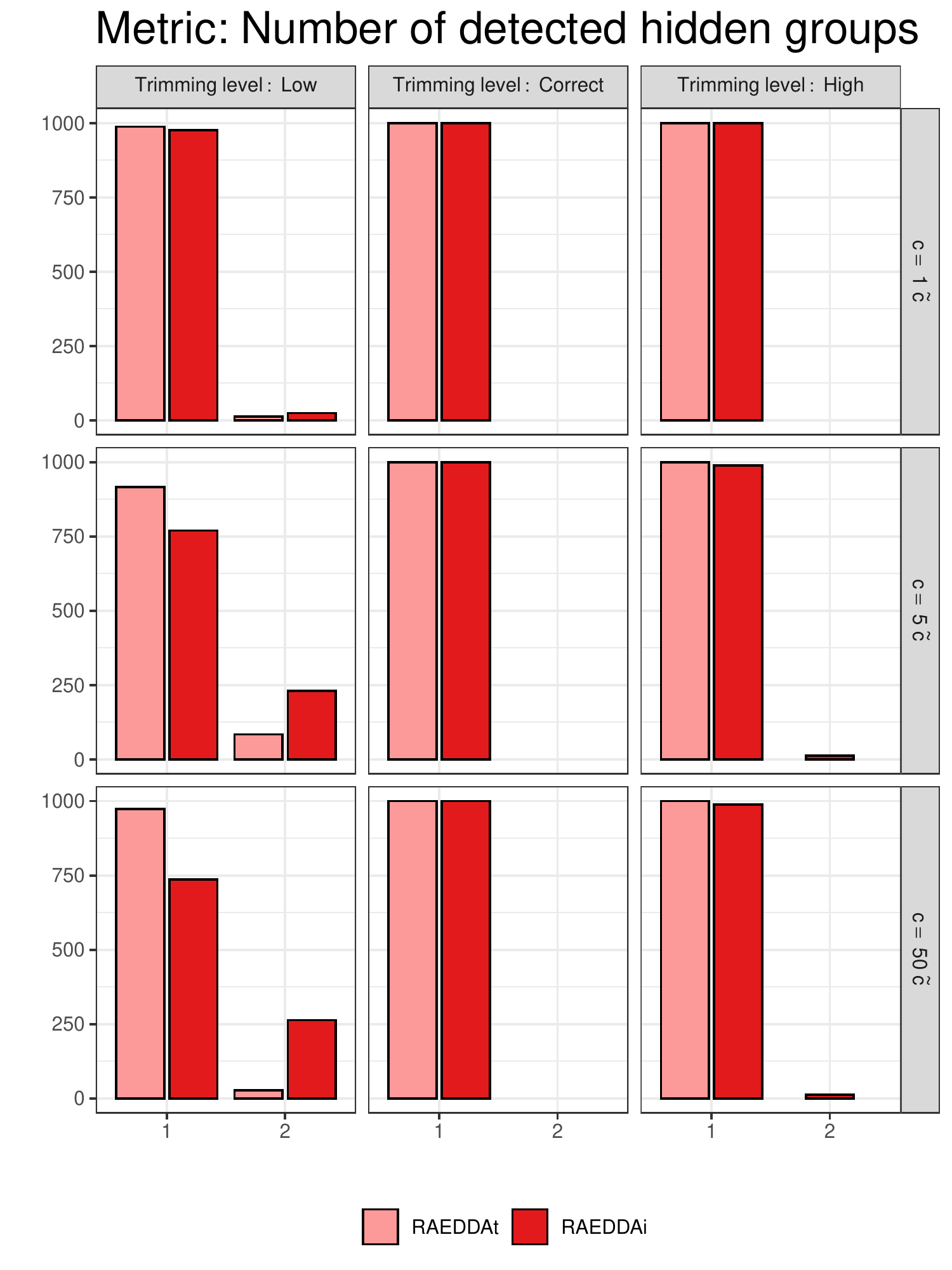}
\end{subfigure}
\caption{Box plots for ARI, \% Hidden Group, \% Label Noise, \% Novelty and number of detected hidden groups for $B = 1000$ Monte Carlo repetitions under different trimming levels and eigenvalue-ratio constraint.}
\label{fig:sens_study_boxplots}
\end{figure*}
The ``\% Label Noise'' metric assesses the ability of our proposal to identify the $Q_l$ incorrectly labelled units in the training set, thus protecting the parameter estimates from bias. Both transductive and inductive approaches perform well regardless of the contamination rate; the number of detected mislabelled units however slightly decreases under the VVV and VVV with overlap simulation scenarios. This is nonetheless due to the more complex covariance structure and to the presence of overlapping groups: this makes the identification of label noise more difficult and less crucial for obtaining reliable inference. The ``\% of Hidden group'' metric in Figure \ref{fig:boxplot_sim_study_RAEDDA_1} shows remarkably good performance in detecting the third unobserved class for the adaptive Discriminant Analysis methods, both for AMDA and its robust generalization RAEDDA. Careful investigation of this peculiar result revealed that the AMDA method tended to merge outlying units and the third (unobserved) class in one single extra group. This effect is intensified for the \texttt{unequal} group proportion, as the sample size of the unobserved class is much smaller than the sizes of the known groups and, when medium and strong contamination is considered, more anomalous units than novelties are present in the test set. Notwithstanding, we notice that the RAEDDA performance in terms of ``\% of Hidden group'' is only slightly lower than its non-robust counterpart: our methodology successfully separates the uniform background noise from the hidden pattern, even when the magnitude of the former is higher than the latter's. Multiple robust initializations are paramount for achieving this result since, as it may be expected, the EM algorithm could be trapped in local maxima due to the discovery of  uninteresting structures within the anomalous units. Even though the AMDA method correctly discovers the presence of an extra class, the associated parameter estimates are completely spoiled by the presence of outliers. Furthermore, the same result does not hold in the two most complex scenarios, where the negative effect of attribute and class noise strongly undermines the adaptive effectiveness of the AMDA model, especially when the transductive estimation is performed. The ``ARI'' metric in Figure \ref{fig:boxplot_sim_study_RAEDDA_2} highlights the predictive power of the RAEDDA model: by means of the MAP rule and the trimming indicator function $\varphi(\cdot)$ the true partition of the test set, that jointly includes known groups, one extra class and the subgroup of $Q_u$ outlying units, is efficiently recovered. As previously mentioned, AMDA fails in separating the uniform noise from the extra Gaussian class, with consequent lower values for the ARI metric. Lastly, the ``\% Novelty'' metric serves the purpose of extending the comparison from the two adaptive models to the novelty detection methods, stemming from the machine learning literature. Particularly, the latter class of algorithms only distinguishes the known patterns (i.e., the first two groups in the training set) and the novelty: in our case the hidden class and the uniform noise. It is evident that, as soon as few noisy data points are added to the training set, both novelty detection methods fail in separating known and novel patterns. In addition, the QDA-ND and SVM-ND performances deteriorate when more complex covariance structures are considered. This unexpected behavior seems due to the fact that 4 out of the 6 dimensions are actually irrelevant for group discrimination and consequent novelty detection, lowering the algorithms performance even under outlier-free scenarios \citep{Evangelista2006,Nguyen2010}.

Notice that the model selection criterion for the RAEDDA method defined in Section \ref{sec:TBIC} was used for identifying not only the number of components but also the parsimonious covariance structure: this always yielded to choose the true parametrization according to the values of $(a,b,c,d,e, f)$. As a last worthy note, the simulation study was performed employing the rationale defined in \eqref{c_tilde} for setting $c$ in the eigenvalue-ratio restriction, whilst the impartial trimming levels $\alpha_l$ and $\alpha_u$ were set high enough to account for the presence of both label noise and outliers. A simple sensitivity study is reported in the upcoming Section for displaying how different choices for the trimming levels and the
eigen-ratio constraint affect the novel procedures.

\subsection{Sensitivity study}
Trimming level and eigenvalue-ratio constraint have a synergic impact on the final solution of robust clustering procedures, 
as shown, for instance, in the extensive simulation study performed by \cite{Coretto2016}. To evaluate their influence in our robust and adaptive classifier, we generate a further $B=1000$ Monte Carlo replications for the EVV covariance structure with \texttt{equal} group proportion and medium contamination level, considering the following combination of hyper-parameters:

\begin{itemize}
\item Trimming levels 
\begin{itemize}
\item Low: $\alpha_l=0.5\times\frac{2Q_l}{N}$, $\alpha_u=0.5\times\frac{Q_u}{M}$
\item Correct: $\alpha_l=\frac{2Q_l}{N}$, $\alpha_u=\frac{Q_u}{M}$
\item High: $\alpha_l=1.5\times\frac{2Q_l}{N}$, $\alpha_u=1.5\times\frac{Q_u}{M}$
\end{itemize}
\item Eigenvalue-ratio constraint
\begin{itemize}
\item Precisely inferred from known groups: $c= \tilde{c}$
\item Slightly larger than known groups: $c= 5\tilde{c}$
\item Considerably larger than known groups: $c= 50\tilde{c}$
\end{itemize}
\end{itemize}
Results for the sensitivity study are displayed in Figure \ref{fig:sens_study_boxplots}, where we report the previously considered metrics. In addition, the right-most graph presents barplots showing the detected number of hidden groups for each repetition, varying trimming levels and eigenvalue-ratio constraint. The ARI, \% Hidden Group, \% Label Noise and \% Novelty metrics are essentially unaffected by the considered $c$ value, whereas the trimming levels have a considerable effect in the classification output. Undoubtedly, underestimating the noise percentage produces far worse results, even though the correct partition is better recovered when the true contamination level is considered. Interestingly, the label noise is almost perfectly detected by setting the ``right'' labeled trimming level, without needing to cautiously overestimate it. The eigenvalue-ratio constraint does have an impact, as expected, when we focus on the appearance of spurious solutions as highlighted in the barplots of Figure \ref{fig:sens_study_boxplots}. Their presence, identified by the incorrect detection of a second hidden group, is positively correlated with the underestimation of the noise level. In particular, notice that $\tilde{c}$ is itself inferred by the estimated covariance matrices for the known groups: that is the reason why some spurious solutions are present even in the plot on the top-left corner. Overall, the inductive approach seems to be more sensitive to the appearance of uninteresting groups.

Even though no extreme situations were found in moving the model hyper-parameters, their correct tuning remains a critical challenge, especially for the trimming levels: a promising idea was recently proposed by \cite{Cerioli2019}, however, further research in the robust classification framework is still to be pursued.
\section{Grapevine microbiome analysis for detection of provenances and varieties} \label{sec:application}
In recent years, the tremendous
advancements in metagenomics
have brought to statisticians a whole new set of questions to be addressed with dedicated methodologies, fostering the fast development of research literature in this field \citep{Waldron2018, Calle2019}. In particular, the role of plant microbiota in grapevine cultivar (\textit{Vitis vinifera L.}) is notably relevant since it has been proven to act as discriminating signature for grape origin and variety \citep{Bokulich2014, Bruni2017a}. Therefore, the employment of microbiome analysis for automatically identifying wine characteristics is a promising approach in the food authenticity domain.

A flexible method that performs online classification of grapevine samples, discriminating potentially fraudulent units from known or previously unseen qualities is likely to have a great impact on the field.

Motivated by two datasets of microbiome composition of grape samples, we validate the performance of the method introduced in Section \ref{sec:RAEDDA} under different contamination and dataset shifts scenarios. 
\begin{figure*}
\includegraphics[scale=.5]{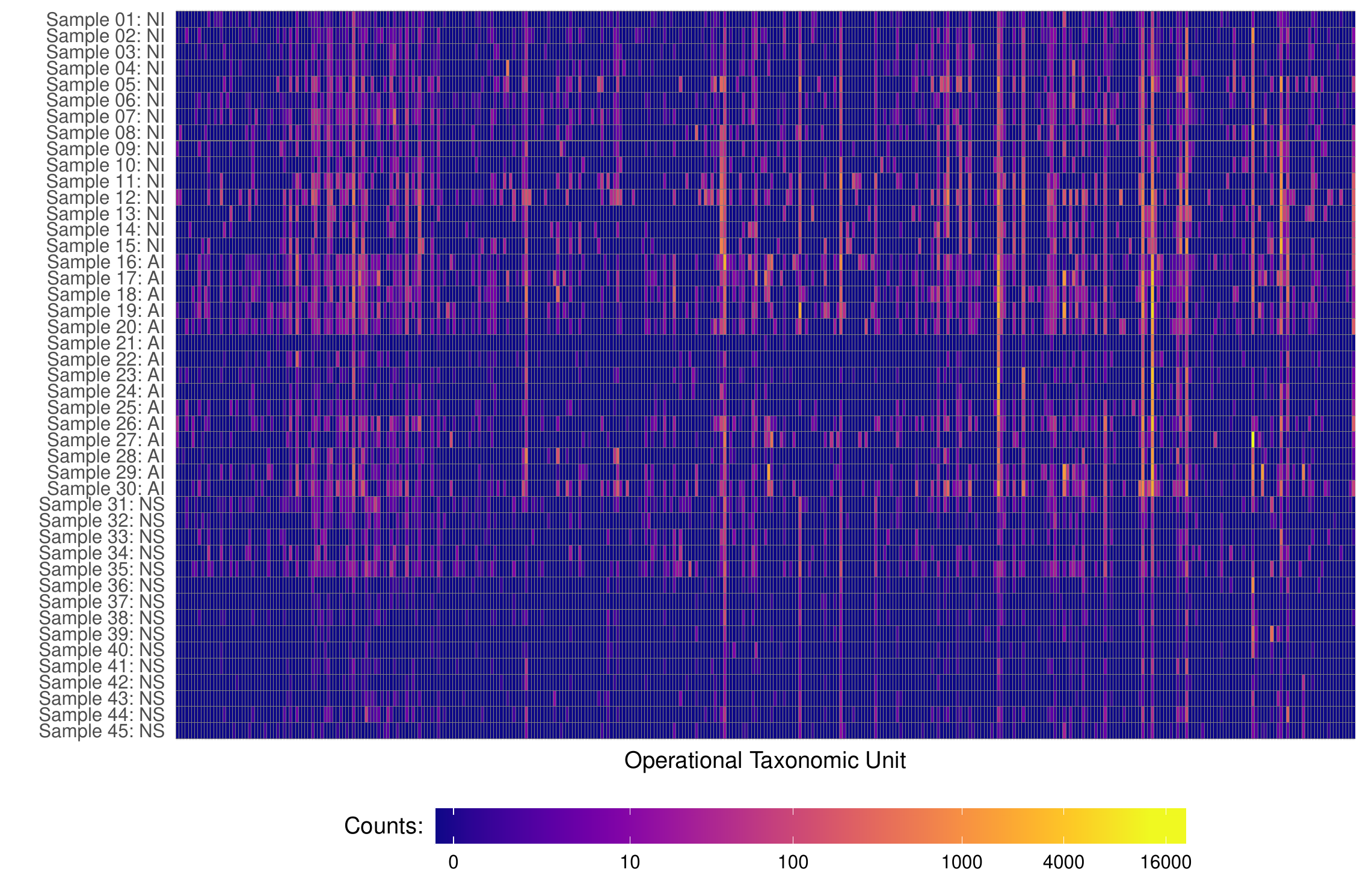}
\centering
\caption{Count table depicting the abundance and distribution of the OTUs resulting from the sequence analysis for each sample in the 3 different regions: Northern Italy (NI), Italian Alps (AI) and Northern Spain (NS). Grape microbiota data.}
\label{fig:OTU}
 \end{figure*}  
\subsection{Grape microbiota of Northern Italy and Spain vineyards} \label{sec:application_1}
\subsubsection{Data}
The first considered dataset reports microbiome composition of $45$ grape samples collected in $3$ different regions having similar pedological features. The first sampling site was the Lombardy Regional Collection in Northern Italy (hereafter NI); the second site was the germplasm collection of E. Mach Foundation in the Trento province, at the foot of the Italian Alps (AI); while the third group of grapes comes from the Government of La Rioja collection, located in Northern Spain (NS). A total of $15$ units were retained from each site. The processes of DNA extraction, sequencing and numbering of microbial composition are thoroughly described in \cite{Mezzasalma2018}: we refer the reader interested in the bioinformatics details to consult that paper and references therein.

At the end of sample preparation, the resulting dataset consists of an abundance table with $836$ features (bacterial communities) defined as Operational Taxonomic Unit (OTU): collapsed clusters of similar DNA sequences that describe the total microbial diversity. For each site, 15 observations are available: a graphical representation of the count table, collapsed at OTU level for ease of visualization, is reported in Figure \ref{fig:OTU}.
\subsubsection{Dimension reduction} \label{sec:preproc_ROBPCA}
Given the high-dimensional nature of the considered dataset and the small sample size, a preprocessing step for reducing the dimensionality is paramount before fitting the RAEDDA model. Focusing on the counting nature of the observations at hand, a natural choice would be to perform probabilistic Poisson PCA (PLNPCA): a flexible methodology based on the Poisson Lognormal model recently introduced in the literature \citep{Apr2018}. Nevertheless, the variational approximation employed for PLNPCA inference makes its generalization from training to test set not so straightforward, and, furthermore, the whole procedure is not robust to outlying observations. Therefore, given the classification framework in which the preprocessing step needs to be embedded, a less domain-specific, yet robust and well-established technique was preferred for dimension reduction.

The considered preprocessing step proceeds as follows: we fit Robust Principal Component Analysis \newline (ROBPCA) to the labelled set, and afterwards we project the test units to the obtained subspace; please refer to \cite{Hubert2005} for a detailed description of the employed methodology. In this way, robust and test-independent (i.e., suitable for either transductive or inductive inference) low-dimensional scores are available for adaptive classification.
 \subsubsection{Anomaly and novelty detection: label noise and one unobserved class} \label{sec:one_anom_lb_application_1}
  \begin{figure}
\includegraphics[scale=.43]{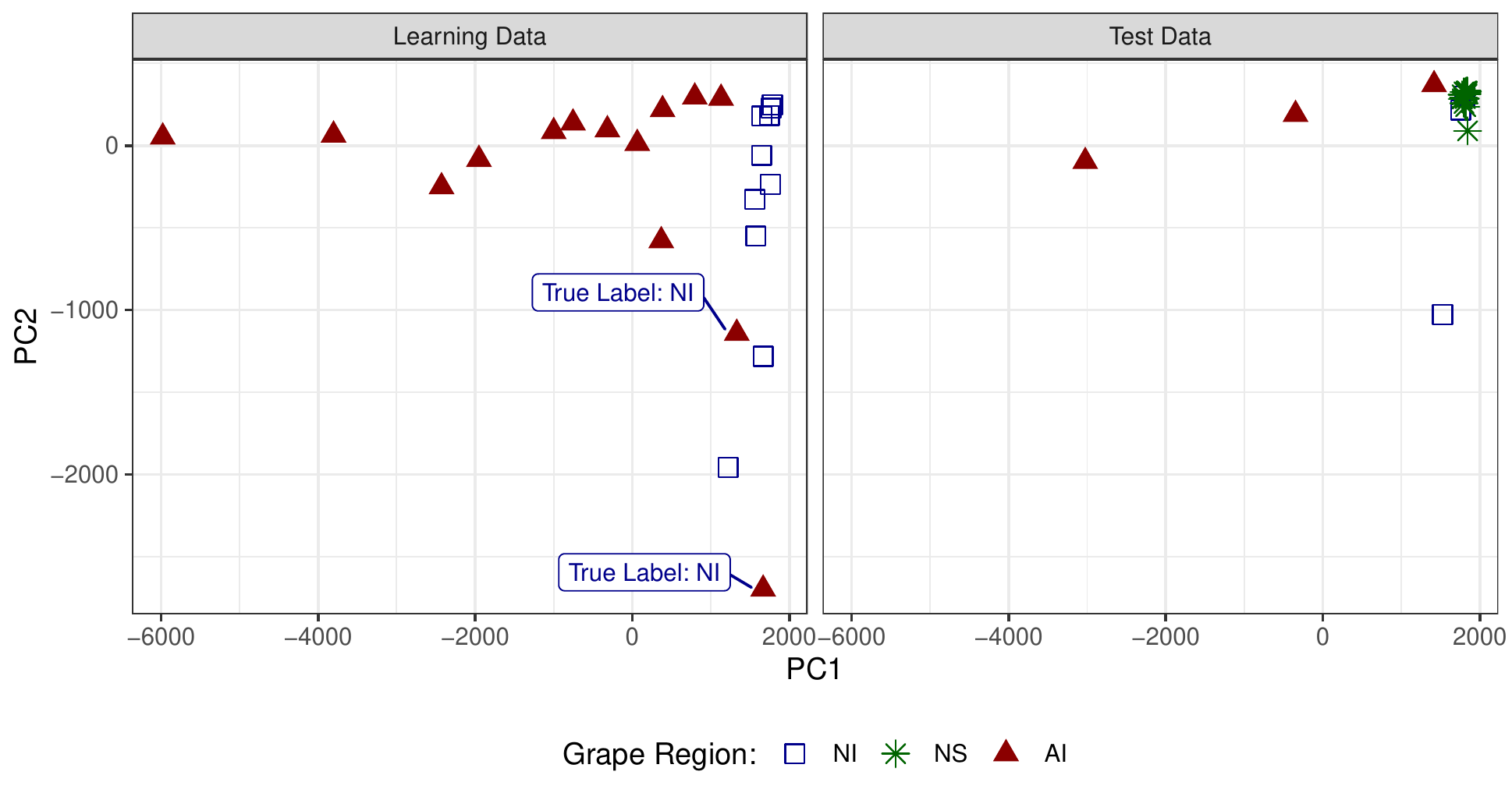}
\centering
\caption{Learning scenario for anomaly and novelty detection of the grapevine microbiota data on the ROBPCA subspace: 1 unobserved region and label noise.}
\label{fig:grape_exp_1}
 \end{figure}
The first experiment involves the random selection of 12 NI and 12 AI units for constructing the training set, with a consequent test set of 21 samples including all the 15 grapes collected in Northern Spain (NS). Furthermore, 2 of the NI units in the learning set are incorrectly labelled as grapes coming from the AI site. The aim of the experiment is therefore to determine whether the RAEDDA method is capable of recovering the unobserved NS class whilst identifying the label noise in the training set. The preprocessing step described in Section \ref{sec:preproc_ROBPCA} is applied prior to perform classification: standard setting for the \texttt{PcaHubert} function in the \texttt{rrcov R} package \citep{Todorov2009a} retains $d=2$ robustly estimated principal components, 
a graphical representation of the learning scenario is reported in Figure \ref{fig:grape_exp_1}.
A RAEDDA model is then employed for building a classification rule, considering both a transductive and an inductive approach. The robust information criterion in \eqref{RBIC} is used for selecting the best patterned structure and, more importantly, the number of extra classes. RBIC values for the two estimation procedures are reported in Tables \ref{tab:RBIC_raeddat_exp1} and \ref{tab:RBIC_raeddai_exp1}: thanks to the orthogonal equivariance of the ROBPCA method,  
we restrict our attention to the subset of diagonal models only. Notice that, in the inductive approach, once the VVI model is selected in the learning phase, only the most flexible diagonal model needs to be fitted to the test data, thanks to the partial order structure of Figure \ref{fig:partial_order_mclust}. Our findings show that the robust information criterion correctly detects the true number of classes $E=3$, in both inferential approaches. 
Regarding anomaly detection, the two units affected by label noise are identified and a posteriori classified as coming from the NI site by the inductive approach. Contrarily, just one out of the two anomalies was captured by the transductive approach. In this and in the  upcoming experiment, trimming levels $\alpha_l=\alpha_u=0.1$ were considered for both training and test sets, while the eigenvalue-ratio restriction was automatically inferred by the estimated group scatters of the known classes.     
 \begin{table*}
\centering
\caption{RBIC for different patterned structures and number of hidden classes for the RAEDDA model, transductive inference. The model with the highest RBIC value is highlighted in bold. Grapevine microbiome data with one unobserved class (NS).}
\label{tab:RBIC_raeddat_exp1}
\begin{tabular}{c|cccccc}
  \noalign{\smallskip}\hline\noalign{\smallskip}
&\multicolumn{6}{|c}{Covariance Structure}\\
\# Classes & EII & VII & EEI & VEI & EVI & \textbf{VVI} \\
 \noalign{\smallskip}\hline\noalign{\smallskip} 
2 & -1278.25 & -1204.55 & -1279.60 & -1208.11 & -1221.39 & -1175.25 \\ 
  \textbf{3} & -1289.24 & -1240.30 & -1291.21 & -1242.67 & -1241.95 & \textbf{-1148.50} \\ 
  4 & -1300.23 & -1254.60 & -1302.20 & -1257.00 & -1256.57 & -1163.34 \\ 
   \noalign{\smallskip}\hline\noalign{\smallskip}
\end{tabular}
\end{table*}
   
 \begin{table*}
 \caption{RBIC for different patterned structures and number of hidden classes for the RAEDDA model, inductive inference. The models with the highest RBIC value are highlighted in bold. Grapevine microbiome data with one unobserved class (NS).}
 \label{tab:RBIC_raeddai_exp1}
 \centering
\begin{tabular}{ c }   
Robust learning phase \\  
\begin{tabular}{c|cccccc}
  \noalign{\smallskip}\hline\noalign{\smallskip}
&\multicolumn{6}{|c}{Covariance Structure}\\
 \# Classes & EII & VII & EEI & VEI & EVI & \textbf{VVI} \\ 
 \noalign{\smallskip}\hline\noalign{\smallskip}
\textbf{2} & -719.26 & -709.13 & -718.97 & -712.11 & -688.40 & \textbf{-678.29} \\ 
   \noalign{\smallskip}\hline\noalign{\smallskip}
\end{tabular} 
\vspace{.5cm}\\
Robust discovery phase \\  
\begin{tabular}{c|c}
 \noalign{\smallskip}\hline\noalign{\smallskip}
&Covariance Structure\\
 \# Classes& \textbf{VVI} \\ 
  \noalign{\smallskip}\hline\noalign{\smallskip}
2 & -642.89 \\ 
  \textbf{3} & \textbf{-509.63} \\ 
  4 & -516.87 \\ 
   \noalign{\smallskip}\hline\noalign{\smallskip}
\end{tabular}\\
\end{tabular}
\end{table*}

Table \ref{tab:grape_exp_1} reports the confusion matrices for the RAEDDA classifier. The model correctly identifies the presence of a hidden class, recovering the true data partition with an accuracy of 86\% (3 misclassified units) and 90\% (2 misclassified units) in the transductive and inductive framework, respectively. 
 \begin{table*}[t]
 \caption{Confusion tables for RAEDDA classifier (transductive and inductive inference) on the test set for the Grapevine microbiome data with one unobserved class (NS).}
 \centering
\begin{tabular}{ cc }   
RAEDDA Transductive & RAEDDA Inductive \\  
\begin{tabular}{ |c|ccc| } 
\hline
&  & Truth & \\
Classification & NI & NS & AI \\ 
  \hline
NI &   1 &   1 &   0 \\ 
  AI &   0 &   0 &   3 \\ 
  HIDDEN GROUP 1 &   2 &  14 &   0 \\
\hline
\end{tabular} &  
\begin{tabular}{ |c|ccc| } 
\hline
&  & Truth & \\
Classification  & NI & NS & AI \\ 
  \hline
NI &   2 &   1 &   0 \\ 
  AI &   0 &   0 &   3 \\ 
  HIDDEN GROUP 1 &   1 &  14 &   0 \\ 
\hline
\end{tabular} \\
\end{tabular}
\label{tab:grape_exp_1}
\end{table*}
Considering the challenging classification problem and the limited sample size, the RAEDDA model shows remarkably good performance. 
\subsubsection{Anomaly and novelty detection: outliers and two unobserved classes}
\begin{figure}[h]
\includegraphics[scale=.43]{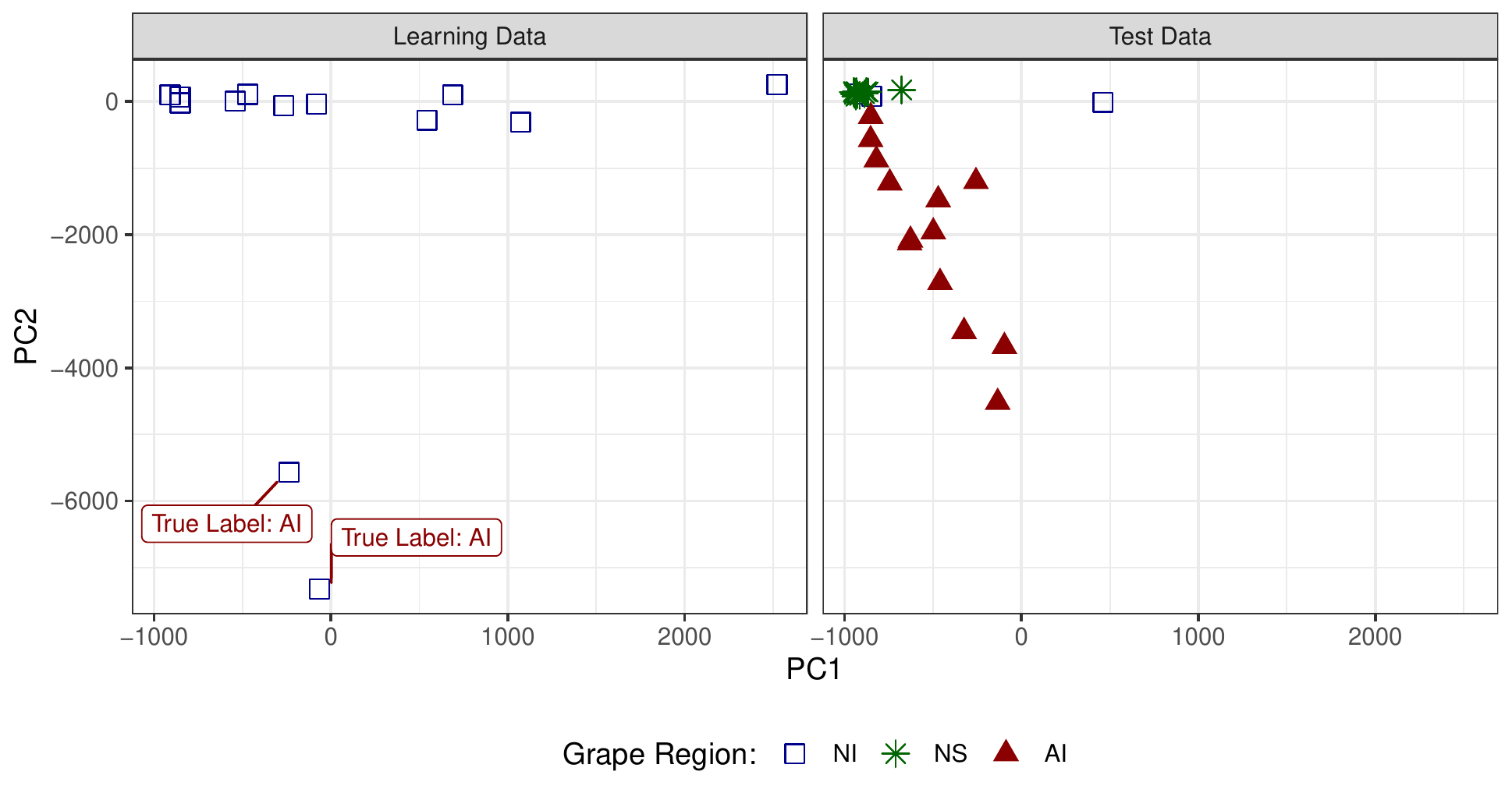}
\centering
\caption{Learning scenario for anomaly and novelty detection of the grapevine microbiota data on the ROBPCA subspace: 2 unobserved regions and outliers in the training set.}
\label{fig:grape_exp_2}
 \end{figure}
This second experiment considers an even more extreme scenario: the training set contains only 14 observations, among which 12 units truly belong to the NI region, while the remaining 2 come from the AI area but with an incorrect NI label. That is, in the remaining 31 unlabelled units there are two sampling sites, namely AI and NS, that need to be discovered.
\begin{table*}
\centering
\caption{RBIC for different patterned structures and number of hidden classes for the RAEDDA model, transductive inference. The model with the highest RBIC value is highlighted in bold. Grapevine microbiome data with two unobserved classes (NS and NI).}
\label{tab:RBIC_raeddat_exp2}
\begin{tabular}{c|cccccc}
  \noalign{\smallskip}\hline\noalign{\smallskip}
&\multicolumn{6}{|c}{Covariance Structure}\\
\# Classes & EII & VII & EEI & VEI & EVI & \textbf{VVI} \\
 \noalign{\smallskip}\hline\noalign{\smallskip} 
1 & -1339.32 & - & -1340.13 & - & - & - \\ 
  2 & -1326.16 & -1251.13 & -1347.46 & -1254.85 & -1351.08 & -1268.71 \\ 
  \textbf{3} & -1337.30 & -1240.35 & -1358.60 & -1244.33 & -1365.89 & \textbf{-1222.04} \\ 
   \noalign{\smallskip}\hline\noalign{\smallskip}
\end{tabular}
\end{table*}
   
 \begin{table*}
 \caption{RBIC for different patterned structures and number of hidden classes for the RAEDDA model, inductive inference. The models with the highest RBIC value are highlighted in bold. Grapevine microbiome data with two unobserved classes (NS and NI).}
 \label{tab:RBIC_raeddai_exp2}
 \centering
\begin{tabular}{ c }   
Robust learning phase \\  
\begin{tabular}{c|cccccc}
  \noalign{\smallskip}\hline\noalign{\smallskip}
&\multicolumn{6}{|c}{Covariance Structure}\\
 \# Classes & EII & VII & \textbf{EEI} & VEI & EVI & VVI \\ 
  \noalign{\smallskip}\hline\noalign{\smallskip}
\textbf{1} & -390.83 & - & \textbf{-364.67 }& - & - & - \\ 
  \noalign{\smallskip}\hline\noalign{\smallskip}
\end{tabular} 
\vspace{.5cm}\\
Robust discovery phase \\  
\begin{tabular}{c|cccc}
 \noalign{\smallskip}\hline\noalign{\smallskip}
&\multicolumn{4}{|c}{Covariance Structure}\\
 \# Classes& EEI & VEI & EVI & \textbf{VVI} \\ 
  \noalign{\smallskip}\hline\noalign{\smallskip}
1 & -3910.27 &  &  &  \\ 
  2 & -1418.22 & -1042.85 & -982.86 & -979.16 \\ 
  \textbf{3} & -1104.38 & -1037.56 & -955.12 & \textbf{-897.46} \\ 
    \noalign{\smallskip}\hline\noalign{\smallskip}
\end{tabular}\\
\end{tabular}
\end{table*}
Likewise in the previous Section, ROBPCA retains $d=2$ principal components when fitted to the training set: the grapevine sample in the robustly estimated subspace is plotted in Figure \ref{fig:grape_exp_2}. Notice in this context the compelling necessity of performing robust dimensional reduction: the two mislabelled observations from the AI area in the training set can be seen as outliers, and a dimensional reduction technique sensitive to them may have introduced masked and/or swamped units. 
The RBIC is used to select the best patterned structure and number of components: results are reported in Tables \ref{tab:RBIC_raeddat_exp2} and \ref{tab:RBIC_raeddai_exp2}.
\begin{table*}[t]
 \caption{Confusion tables for RAEDDA classifier (transductive and inductive inference) on the test set for the Grapevine microbiome data with two unobserved classes (NS and NI).}
 \centering
\begin{tabular}{ cc }   
RAEDDA Transductive & RAEDDA Inductive \\  
\begin{tabular}{ |c|ccc| } 
\hline
&  & Truth & \\
Classification & NI & NS & AI \\ 
  \hline
NI &   1 &   1 &   1 \\ 
  HIDDEN GROUP 1 &   0 &   0 &  12 \\ 
  HIDDEN GROUP 2 &   2 &  14 &   0 \\ 
\hline
\end{tabular} &  
\begin{tabular}{ |c|ccc| } 
\hline
&  & Truth & \\
Classification  & NI & NS & AI \\ 
  \hline
NI &   1 &   0 &   1 \\ 
  HIDDEN GROUP 1 &   2 &  15 &   0 \\ 
  HIDDEN GROUP 2 &   0 &   0 &  12 \\ 
\hline
\end{tabular} \\
\end{tabular}
\label{tab:grape_exp_2}
\end{table*}
Again, also in this more extreme experiment both inferential procedures recover the true number of sites from which the grapes were sampled. Due to the ROBPCA output, in both transductive and inductive approaches the wrongly labelled units in the training set are easily trimmed off and identified as belonging to an area different from NI. Classification results for the chosen model are reported in Table \ref{tab:grape_exp_2}, where the recovered data partition notably agrees with the 3 different sampling sites, with only 4 and 3 misclassified units for the transductive and inductive estimation, respectively.
\subsection{Must microbiota of Napa and Sonoma Counties, California}
\subsubsection{Data}
The second dataset reports microbiome composition of $239$ crushed grapes (must) for $3$ different wine varieties; namely Dolce, Cabernet Sauvignon and Chardonnay grown throughout Napa and Sonoma Counties, California. The considered samples are a subset of the ``Bokulich Microbial Terroir'' study, and are publicly available in the QIITA database under accession no. 10119 \newline (http://qiita.ucsd.edu/study/description/10119). Likewise for the previous analysis, technical details concerning the retrieval of the final abundance table are deferred to the original paper \citep{Bokulich2016}. Ultimately, sample features encompass the counts of $9943$ bacterial communities (OTU) and data partition with respect to wine type is as follows: $99$ must units belong to Cabernet Sauvignon, $114$ to Chardonnay and $26$ to Dolce variety. 

\subsubsection{Dimension reduction} \label{sec:data_reduction_bokulich}
The high-dimensional nature of the problem requires a feature reduction technique  to be performed prior to employ our anomaly and novelty detection method. Given that the number of bacterial communities is almost $12$ times larger in magnitude with respect to the dataset of Section \ref{sec:application_1}, a more standard microbiome preprocessing procedure has been adopted. By means of the QIIME2 bioinformatics platform \citep{Bolyen2019}, Bray-Curtis dissimilarity metrics (evenly sampled at $2000$ reads per sample) are computed between each pair of units. From the resulting distance matrix, a robust version of the Principal Coordinates Analysis  (PCoA) is performed considering a robust singular value decomposition \citep{hawkins2001robust} within the classical multidimensional scaling algorithm. Lastly, a total of $p=10$ coordinates are retained for the subsequent study. Notice that, as highlighted in Section 5 of \cite{hawkins2001robust}, the eigenvectors returned by the robust singular value decomposition are, in general, not orthogonal. Therefore, differently from the previous application, the whole set of 14 covariance structures will be considered when fitting the RAEDDA models.

\begin{table*}
\centering
\caption{RBIC for different patterned structures and number of hidden classes for the RAEDDA model, transductive inference. The model with the highest RBIC value is highlighted in bold. Must microbiota data with one unobserved class (Dolce).}
\label{tab:RBIC_raeddat_data_2}
\begin{tabular}{c|ccccccc}
  \noalign{\smallskip}\hline\noalign{\smallskip}
&\multicolumn{7}{|c}{Covariance Structure}\\
\# Classes & EII & VII & EEI & VEI & EVI & VVI & EEE \\ 
  \noalign{\smallskip}\hline\noalign{\smallskip}
2 & 4881.67 & 5984.40 & 5927.50 & 6617.88 & 6106.50 & 6966.14 & 6845.79 \\ 
  3 & 5432.89 & 6278.19 & 6404.65 & 7020.50 & 6553.41 & 7180.49 & 7296.24 \\ 
  4 & 5479.22 & 6325.52 & 6456.40 & 7037.37 & 6595.53 & 7228.00 & 7502.04 \\ 
   \noalign{\medskip}\hline\noalign{\medskip}
   \# Classes & EVE & VEE & VVE & EEV & \textbf{VEV} & EVV & VVV \\ 
2 & 7277.23 & 7460.51 & 7656.49 & 7443.72 & 8040.17 & 7233.28 & 8067.80 \\ 
 \textbf{3} &  & 7760.96 & 8281.65 & 7985.32 & \textbf{8464.17} & 7895.36 & 8400.12 \\ 
  4 &  & 7934.52 &  & 7875.29 & 8338.43 &  & 8203.00 \\
      \noalign{\smallskip}\hline\noalign{\smallskip}
\end{tabular}
\end{table*}
   
 \begin{table*}
 \caption{RBIC for different patterned structures and number of hidden classes for the RAEDDA model, inductive inference. The models with the highest RBIC value are highlighted in bold. Must microbiota data with one unobserved class (Dolce).}
 \label{tab:RBIC_raeddai_data_2}
 \centering
\begin{tabular}{ c }   
Robust learning phase \\  
\begin{tabular}{c|ccccccc}
  \noalign{\smallskip}\hline\noalign{\smallskip}
&\multicolumn{7}{|c}{Covariance Structure}\\
\# Classes & EII & VII & EEI & VEI & EVI & VVI & EEE \\ 
 \noalign{\smallskip}\hline\noalign{\smallskip}
2 & 3419.46 & 3883.43 & 4219.49 & 4521.48 & 4289.74 & 4642.67 & 4997.97 \\ 
   \noalign{\medskip}\hline\noalign{\medskip}
   \# Classes & EVE & VEE & VVE & EEV & \textbf{VEV} & EVV & VVV \\ 
 \noalign{\smallskip}\hline\noalign{\smallskip}
\textbf{2} & 5219.08 & 5158.22 & 5385.09 & 5281.61 & \textbf{5532.97}& 5258.35 & 5512.86 \\ 
 \noalign{\smallskip}\hline\noalign{\smallskip}
\end{tabular} 
\vspace{.5cm}\\
Robust discovery phase \\  
\begin{tabular}{c|cc}
 \noalign{\smallskip}\hline\noalign{\smallskip}
&Covariance Structure\\
 \# Classes& \textbf{VEV} & VVV \\ 
\noalign{\smallskip}\hline\noalign{\smallskip}
2 & -8496.47 &  \\ 
  \textbf{3} & \textbf{3020.84} & 3007.11 \\ 
  4 & 3013.27 & 2904.01 \\ 
   \noalign{\smallskip}\hline\noalign{\smallskip}
\end{tabular}\\
\end{tabular}
\end{table*}
 \subsubsection{Anomaly and novelty detection: label noise and one unobserved class}

The dataset is randomly partitioned in labeled and unlabeled sets: the former is composed by $80$ Chardonnay and $
70$ Cabernet Sauvignon units, while the latter by $24$ Chardonnay, $29$ Cabernet Sauvignon and $22$ Dolce units. The remaining $4$ samples from the Dolce variety are appended to the learning set wrongly setting their label to be Cabernet Sauvignon. The adulteration procedure mimics the one in Section \ref{sec:one_anom_lb_application_1}, nevertheless, this second dataset poses a more challenging problem due to the higher feature dimension, even after its reduction according to the procedure described in Section \ref{sec:data_reduction_bokulich},  and the small sample size of the unobserved class. The robust information criterion in \eqref{RBIC} is employed for selecting the best patterned structure and the number of extra classes: Tables \ref{tab:RBIC_raeddat_data_2} and \ref{tab:RBIC_raeddai_data_2} report its value under transductive and inductive inference, respectively. The hidden class is correctly discovered by both approaches, as it can be seen in the confusion matrices of Table \ref{tab:must_exp}. Trimming levels $\alpha_l=0.03$ and $\alpha_u=0.05$ are sufficient for identifying the units with label noise and correctly assigning them to the newly revealed class in the robust discovery phase (inductive inference). As expected, the overall classification accuracy is lower in this dataset with respect to the previous one, this is mostly driven by the difficulty in discriminating Chardonnay and Cabernet Sauvignon musts.

\begin{table*}[t]
 \caption{Confusion tables for RAEDDA classifier (transductive and inductive inference) on the test set for the must microbiota data with one unobserved class (Dolce).}
 \centering
\begin{footnotesize}
\begin{tabular}{ cc }   
RAEDDA Transductive & RAEDDA Inductive \\  
\begin{tabular}{ |c|ccc| } 
\hline
&  & Truth & \\
Classification & Cabernet S. & Chardonnay & Dolce \\ 
  \hline
Cabernet S. &  24 &  11 &   0 \\ 
  Chardonnay &   2 &  22 &   0 \\ 
  HIDDEN GROUP 1 &   3 &   1 &  22 \\ 
\hline
\end{tabular} & 
\begin{tabular}{ |c|ccc| } 
\hline
&  & Truth & \\
Classification & Cabernet S. & Chardonnay & Dolce \\ 
  \hline
Cabernet S. &  25 &   9 &   0 \\ 
  Chardonnay &   1 &  25 &   0 \\ 
  HIDDEN GROUP 1 &   3 &   0 &  22 \\ 
\hline
\end{tabular} \\
\end{tabular}
\end{footnotesize}
\label{tab:must_exp}
\end{table*}

All in all, considering these and additional experiments not reported in the present paper, the inductive approach seems to perform slightly better in terms of anomaly and novelty detection, especially if the sample size of the hidden classes is small. This had already been noted in \cite{Bouveyron}, and it may be even more evident in our proposal due to the augmented test set (see end of Section \ref{sec:EM_inductive_l}) employed in the discovery phase. For instance, in this experiment, the four Dolce units that are trimmed off in the learning phase come back again in the parameter estimation of the discovery phase, improving the classifier efficiency. Contrarily, the transductive approach simply does not account for them when estimating the parameters of the Dolce group.

Even though domain-expert supervision will always be crucial for class interpretation when extra groups are detected, an automatic pipeline that performs microbiome composition, dimension reduction and robust and adaptive classification seems a promising procedure for enhancing the quality, speed and mechanization of food authenticity analyses.

\section{Concluding Remarks} \label{sec:conclusion}
In the present paper we have proposed a model-based discriminant analysis method for anomaly and novelty detection. We have shown that the methodology effectively performs classification in presence of label noise, outliers and unobserved classes in the test set. By incorporating impartial trimming and eigenvalue-ratio constraints, our proposal robustly estimates model parameters of known and hidden classes, identifying as a by-product wrongly labelled and/or adulterated observations. Considering a parsimonious family of patterned models, two flexible EM-based approaches have been proposed for parameter estimation: one based on the union of training and test sets, and the other made of two phases, performing sequential inference for known and hidden groups. Furthermore, we let the latter approach exploit the partial order structure of the parsimonious models, deriving fast and closed-form solutions for estimating the parameters of the extra classes. The resulting methodology includes several model-based classification methods as special cases. A robust data-driven criterion has been adapted for selecting the number of unobserved groups and constraint strength in covariances estimation. An extensive simulation study and applications on two grapevine microbiome datasets have proved the effectiveness of our proposal. Particularly, the classifier capability in discriminating (known and previously unobserved) grape provenances and varieties, within an adulterated context, may lead to promising developments in the food authenticity domain.

Further research directions include a data-driven procedure for selecting reasonable values for the trimming levels, and a metric that automatically categorizes trimmed units as being affected by label and/or attribute noise. Additionally, the definition of a general framework for robust and adaptive variable selection and classification, suitable for data of large dimensions, is imperative in domains like chemometrics, computer vision and genetics: a proposal is currently under study and it will be object of future developments.

\subsubsection*{Acknowledgements}
The authors are grateful to Anna Sandionigi, Lorenzo Guzzetti, Maurizio Casiraghi  and
Massimo Labra for fruitful discussions and domain-knowledge sharing for our Grapevine microbiome analyses for detection of provenances and varieties. In particular, authors thank Anna Sandionigi for her decisive help in performing the routines described in Section \ref{sec:data_reduction_bokulich} and for her support throughout the must samples analysis. We also would like to thank the Editor, Associate Editor and Referees whose suggestions and comments enhanced the quality of the paper.
Brendan Murphy's work was supported by the Science Foundation Ireland Insight Research Centre grant (12/RC/2289\_P2). 
  
\section*{Appendix A: Inductive covariance matrices estimation}
This appendix provides closed form solutions for the estimation of the covariance matrices $\boldsymbol{\Sigma}_h$, $ h=G+1,\ldots,E$ of the unobserved classes via the inductive approach; our main reference here is the seminal paper of \cite{Celeux1995}, where patterned covariance matrices were firstly defined and algorithms for their ML estimation were proposed. In the robust discovery phase only the parameters for the $H=E-G$ densities need to be estimated, according to the available patterned models, given the one considered in the Learning Phase (see Figure \ref{fig:partial_order_mclust}). Denote with $\boldsymbol{W}_h=\sum_{m=1}^{M^{*}} \varphi(\mathbf{y}^{*}_m)\hat{z}^{*}_{mh}\left[ \left(\mathbf{y}^{*}_{m}-\hat{\boldsymbol{\mu}}_{h}\right)\left(\mathbf{y}^{*}_{m}-\hat{\boldsymbol{\mu}}_{h}\right)^{\prime}\right]$ and let $\boldsymbol{W}_h=\boldsymbol{L}_h\boldsymbol{\Delta}_h\boldsymbol{L}^{'}_h$ be its eigenvalue decomposition. Further, consider $n_h=\sum_{m=1}^{M^{*}} \varphi(\mathbf{y}^{*}_m)\hat{z}^{*}_{mh}$ for $h=G+1,\ldots, E$. Lastly, denote with a bar the estimates obtained in the robust learning phase for the $G$ known groups: they are fixed and should not be changed. The formulae needed for the parameter updates are as follows:
\begin{itemize}  
\item VII model: $\boldsymbol{\Sigma}_h=\lambda_h \boldsymbol{I}$
\begin{equation*}
\hat{\lambda}_h= \frac{\hbox{tr}(\boldsymbol{W}_h)}{p \: n_h}, \qquad h=G+1,\ldots,E.
\end{equation*}

\item VEI model: $\boldsymbol{\Sigma}_h=\lambda_h \bar{\boldsymbol{A}}$
\begin{equation*}
\hat{\lambda}_h= \frac{\hbox{tr}(\boldsymbol{W}_h \bar{A}^{-1})}{p \: n_h}, \qquad h=G+1,\ldots,E.
\end{equation*}

\item EVI model: $\boldsymbol{\Sigma}_h=\bar{\lambda} \boldsymbol{A}_h$
\begin{equation*}
\hat{\boldsymbol{A}}_h= \frac{\hbox{diag}(\boldsymbol{W}_h)}{|\hbox{diag}(\boldsymbol{W}_h)|^{1/p}}, \qquad h=G+1,\ldots,E.
\end{equation*}

\item VVI model: $\boldsymbol{\Sigma}_h=\lambda_h \boldsymbol{A}_h$
\begin{equation*}
\hat{\lambda}_h= \frac{|\hbox{diag}(\boldsymbol{W}_h)|^{1/p}}{n_h}, \qquad h=G+1,\ldots,E.
\end{equation*}
\begin{equation*}
\hat{\boldsymbol{A}}_h= \frac{\hbox{diag}(\boldsymbol{W}_h)}{|\hbox{diag}(\boldsymbol{W}_h)|^{1/p}}, \qquad h=G+1,\ldots,E.
\end{equation*}

\item VEE model: $\boldsymbol{\Sigma}_h=\lambda_h \bar{\boldsymbol{D}}\bar{\boldsymbol{A}}\bar{\boldsymbol{D}}^{'}$

Let $\bar{\boldsymbol{C}}=\bar{\boldsymbol{D}}\bar{\boldsymbol{A}}\bar{\boldsymbol{D}}^{'}$ and
\begin{equation*}
\hat{\lambda}_h= \frac{\hbox{tr}(\boldsymbol{W}_h \bar{\boldsymbol{C}}^{-1})}{p \: n_h}, \qquad h=G+1,\ldots,E. 
\end{equation*}

\item EVE model: $\boldsymbol{\Sigma}_h=\bar{\lambda} \bar{\boldsymbol{D}}\boldsymbol{A}_h\bar{\boldsymbol{D}}^{'}$
\begin{equation*}
\hat{\boldsymbol{A}}_h= \frac{\hbox{diag}(\bar{\boldsymbol{D}}^{'}\boldsymbol{W}_h\bar{\boldsymbol{D}})}{|\hbox{diag}(\bar{\boldsymbol{D}}^{'}\boldsymbol{W}_h\bar{\boldsymbol{D}})|^{1/p}}, \qquad h=G+1,\ldots,E.
\end{equation*}
\item EEV model: $\boldsymbol{\Sigma}_h=\bar{\lambda} \boldsymbol{D}_h\bar{\boldsymbol{A}}\boldsymbol{D}_h^{'}$
\begin{equation*}
\hat{\boldsymbol{D}}_h= \boldsymbol{L}_h, \qquad h=G+1,\ldots,E.
\end{equation*}
\item VVE model: $\boldsymbol{\Sigma}_h=\lambda_h \bar{\boldsymbol{D}}\boldsymbol{A}_h\bar{\boldsymbol{D}}{'}$

Let $\boldsymbol{R}_h=\lambda_h \boldsymbol{A}_h$
\begin{equation*}
\hat{\boldsymbol{R}}_h= \frac{1}{n_h}\hbox{diag}(\bar{\boldsymbol{D}}^{'}\boldsymbol{W}_h\bar{\boldsymbol{D}}), \qquad h=G+1,\ldots,E.
\end{equation*}
and, subsequently
\begin{equation*}
\hat{\lambda}_h= |\hat{\boldsymbol{R}}_h|^{1/p}, \qquad h=G+1,\ldots,E.
\end{equation*}
\begin{equation*}
\hat{\boldsymbol{A}}_h= \frac{1}{\hat{\lambda}_h}\hat{\boldsymbol{R}}_h, \qquad h=G+1,\ldots,E.
\end{equation*}
\item VEV model: $\boldsymbol{\Sigma}_h=\lambda_h \boldsymbol{D}_h\bar{\boldsymbol{A}}\boldsymbol{D}_h^{'}$

\begin{equation*}
\hat{\boldsymbol{D}}_h= \boldsymbol{L}_h, \qquad h=G+1,\ldots,E.
\end{equation*}

\begin{equation*}
\hat{\lambda}_h= \frac{\hbox{tr}(\boldsymbol{W}_h \hat{\boldsymbol{D}}_h \bar{\boldsymbol{A}}^{-1}\hat{\boldsymbol{D}}_h{'})}{p \: n_h}, \qquad h=G+1,\ldots,E. 
\end{equation*}

\item EVV model: $\boldsymbol{\Sigma}_h=\bar{\lambda} \boldsymbol{D}_h\boldsymbol{A}_h\boldsymbol{D}_h^{'}$

Let $\boldsymbol{C}_h= \boldsymbol{D}_h\boldsymbol{A}_h\boldsymbol{D}_h^{'}$
\begin{equation*}
\hat{\boldsymbol{C}}_h= \frac{\boldsymbol{W}_h}{|\boldsymbol{W}_h|^{1/p}}, \qquad h=G+1,\ldots,E.
\end{equation*}
$\hat{\boldsymbol{A}}_h$, $\hat{\boldsymbol{D}_h}$ are obtained through the eigenvalue decomposition of $\hat{\boldsymbol{C}}_h$, $h=G+1,\ldots,E$.
\item VVV model: $\boldsymbol{\Sigma}_h=\lambda_h \boldsymbol{D}_h\boldsymbol{A}_h\boldsymbol{D}_h^{'}$
\begin{equation*}
\hat{\boldsymbol{\Sigma}}_h=\frac{1}{n_h}\boldsymbol{W}_h
\end{equation*}
$\hat{\lambda_h}$, $\hat{\boldsymbol{A}}_h$, $\hat{\boldsymbol{D}_h}$ are obtained through the eigenvalue decomposition of $\hat{\boldsymbol{\Sigma}}_h$, $h=G+1,\ldots,E$.
\end{itemize}
Lastly, it is easy to see that whenever the model in the discovery phase is EII, EEI or EEE, no extra parameters need to be estimated for the covariance matrices of the hidden groups.

\end{document}